\documentclass[journal]{IEEEtran}
\usepackage{cite}
\usepackage{epstopdf}
\usepackage{amsfonts,dsfont,amsthm,amsmath,amssymb,amscd,epsfig,bm,subfigure,graphicx,mathtools}
\usepackage{textcomp}
\usepackage{color,soul}
\usepackage[utf8]{inputenc}
\usepackage[T1]{fontenc}
\usepackage{tabularx} 
\usepackage{multirow}
\usepackage{wasysym}
\usepackage{algorithm}
\usepackage{algorithmic}
\usepackage{suffix}
\usepackage{url}
\usepackage{multicol}
\usepackage{verbatim}
\usepackage{acro} \acsetup{foreign-format = {\emph}}
\title{Neurosciences and 6G: Lessons from\\ and Needs of Communicative Brains}
\author{
Renan C. Moioli, 
Pedro H. J. Nardelli,~\IEEEmembership{Senior Member,~IEEE,}
Michael Taynnan Barros, 
Walid Saad,~\IEEEmembership{Fellow,~IEEE,}
Amin Hekmatmanesh,
Pedro Gória, 
Arthur S. de Sena,~\IEEEmembership{Student Member,~IEEE,} 
Merim Dzaferagic,
Harun Siljak,~\IEEEmembership{Member,~IEEE,}
Werner van Leekwijck,
Dick Carrillo,~\IEEEmembership{Member,~IEEE,}
Steven Latré
\thanks{RCM is with Federal University of Rio Grande do Norte, Brazil; MTB is with Tampere University, Finland; AH, PHJN, AS, and DC are with LUT University, Finland; WS is with Virginia Tech, USA; PG is with Instituto Nacional de Telecomunicações, Brazil; MD and HS are with Trinity College Dublin, Ireland; WL and SL are with Antwerpen University, Belgium. 
This paper is partly supported by Academy of Finland via: (a) ee-IoT project n.319009, (b) FIREMAN consortium CHIST-ERA/n.326270, and (c) EnergyNet Research Fellowship n.321265/n.328869.
Corresponding author (PHJN): pedro.nardelli@lut.fi}
}
\DeclareAcronym{ai}{
	short = AI,
	long  = Artificial Intelligence,
	class = acrs,
}
\DeclareAcronym{6g}{
	short = 6G,
	long  = sixth generation of mobile networks,
	class = acrs,
}
\DeclareAcronym{bmi}{
	short = BMI,
	long  = Brain-Machine Interface,
	class = acrs,
}
\DeclareAcronym{btc}{
	short = BTC,
	long  = Brain-Type Communications,
	class = acrs,
}
\DeclareAcronym{qope}{
	short = QoPE,
	long  = Quality-of Physical-Experience,
	class = acrs,
}
\DeclareAcronym{iobnt}{
	short = IoBNT,
	long  = Internet-of-Bio-Nano-Things,
	class = acrs,
}
\DeclareAcronym{bcv}{
	short = BCV,
	long  = Brain-Controlled Vehicles,
	class = acrs,
}
\DeclareAcronym{ic}{
	short = IC,
	long  = integrated circuit,
	class = acrs,
}
\DeclareAcronym{lfp}{
	short = LFP,
	long  = local field potentials,
	class = acrs,
}
\DeclareAcronym{htc}{
	short = HTC,
	long  = Human-Type Communication,
	class = acrs,
}
\DeclareAcronym{iot}{
	short = IoT,
	long  = Internet of Things,
	class = acrs,
}
\DeclareAcronym{mtc}{
	short = MTC,
	long  = Machine-Type Communication,
	class = acrs,
}
\DeclareAcronym{xr}{
	short = XR,
	long  = Extended Reality,
	class = acrs,
}
\DeclareAcronym{b2b}{
	short = B2B,
	long  = Brain-to-Brain,
	class = acrs,
}
\DeclareAcronym{d2d}{
	short = D2D,
	long  = Device-to-Device,
	class = acrs,
}
\DeclareAcronym{qos}{
	short = QoS,
	long  = Quality-of-Service,
	class = acrs,
}
\DeclareAcronym{qoe}{
	short = QoE,
	long  = Quality-of-Experience,
	class = acrs,
}
\DeclareAcronym{embb}{
	short = eMBB,
	long  = enhanced mobile broadband,
	class = acrs,
}
\DeclareAcronym{urllc}{
	short = URLLC,
	long  = ultra reliable low latency communication,
	class = acrs,
}
\DeclareAcronym{mmtc}{
	short = mMTC,
	long  = massive Machine-Type Communication,
	class = acrs,
}
\DeclareAcronym{aoi}{
	short = AoI,
	long  = age of information,
	class = acrs,
}
\DeclareAcronym{voi}{
	short = VoI,
	long  = value of information,
	class = acrs,
}
\DeclareAcronym{mimo}{
	short = MIMO,
	long  = multiple input multiple output,
	class = acrs,
}
\DeclareAcronym{snr}{
	short = SNR,
	long  = signal-to-noise ratio,
	class = acrs,
}
\DeclareAcronym{irs}{
	short = IRS,
	long  = intelligent reflecting surface,
	class = acrs,
}
\DeclareAcronym{wsn}{
	short = WSN,
	long  = Wireless Sensor Networks,
	class = acrs,
}
\DeclareAcronym{fft}{
	short = FFT,
	long  = Fast  Fourier  Transform,
	class = acrs,
}
\DeclareAcronym{isn}{
	short = ISN,
	long  = intelligent sensor network,
	class = acrs,
}
\DeclareAcronym{ann}{
	short = ANN,
	long  = artificial neural network,
	class = acrs,
}
\DeclareAcronym{snn}{
	short = SNN,
	long  = Spiking Neural Network,
	class = acrs,
}
\DeclareAcronym{stdp}{
	short = STDP,
	long  = spike-timing-dependent plasticity,
	class = acrs,
}
\DeclareAcronym{htm}{
	short = HTM,
	long  = hierarchical temporal memory,
	class = acrs,
}
\DeclareAcronym{ar}{
	short = AR,
	long  = Augmented Reality,
	class = acrs,
}

\DeclareAcronym{dcsk}{
	short = DCSK,
	long  =  differential chaos-shift-keying,
	class = acrs,
}

\DeclareAcronym{bpsk}{
	short = BPSK,
	long  = binary phase-shift keying,
	class = acrs,
}
\DeclareAcronym{lpi}{
	short = LPI,
	long  = low probability of intercept,
	class = acrs,
}
\DeclareAcronym{pn}{
	short = PN,
	long  = pseudo-noise,
	class = acrs,
}
\begin{document}

\maketitle

\begin{abstract}
This paper presents the first comprehensive tutorial on a promising research field located at the frontier of two well-established domains: Neurosciences and wireless communications, motivated by the ongoing efforts to define how the \ac{6g} will be. In particular, this tutorial first provides a novel integrative approach that bridges the gap between these two, seemingly disparate fields. Then, we present the state-of-the-art and key challenges of these two topics.
In particular, we propose a novel systematization that divides the contributions into two groups, one focused on what neurosciences will offer to \acs{6g} in terms of new applications and systems architecture  (\textit{Neurosciences for Wireless}), and the other focused on how wireless communication theory and \acs{6g} systems can provide new ways to study the brain (\textit{Wireless for Neurosciences}). 
For the first group, we concretely explain how current scientific understanding of the brain would enable new application for \acs{6g} within the context of a new type of service that we dub \textit{brain-type communications} and that has more stringent requirements than human- and machine-type communication. In this regard, we expose the key requirements of brain-type communication services and we discuss how future wireless networks can be equipped to deal with such services. Meanwhile, for the second group, we thoroughly explore modern communication system paradigms, including Internet of Bio-nano Things and chaos-based communications, in addition to highlighting how complex systems tools can help bridging \acs{6g} and neuroscience applications.
Brain-controlled vehicles are then presented as our case study to demonstrate for both groups the potential created by the convergence of neurosciences and wireless communications in \acs{6g}. 
All in all, this tutorial is expected to provide a largely missing articulation between these two emerging fields while delineating concrete ways to move forward in such an interdisciplinary endeavor. 
\end{abstract}


\section{Introduction}
%

The last two decades witnessed tremendous new developments in information and communication technologies, the most remarkable of which being recent advances in  wireless communications and \ac{ai}.
At the same time, the scientific understanding of the nervous system and the brain has also substantially grown. In fact, brain research is seen as arguably the most anticipated field of research for the coming decade.
This was not a historical coincidence: the evolution of both domains are strongly interlinked. 
For example, on the one hand, the steep growth rates of technological advances in sensors, digital processing, and computational models have always supported the research in neurosciences while, on the other hand, the knowledge of how neurons and the neurological system work supported the development of computational methods based on \ac{ann}~\cite{ML00}.
An interesting interview about the topic can be found in \cite{strukov2019building}.

Neurosciences and \ac{6g} are converging in the context of several recent wireless and \ac{ai} developments where both are going to the edge: wireless is quickly heading towards nano-communication while \ac{ai} is moving towards edge intelligence at the sensor itself based on neuromorphic computing and various edge \ac{ai} techniques such as federated learning~\cite{bonawitz2019federated,FEDSURV00,FLL00,ARI00,MINGBIP}.
Futuristic technological solutions like Neuralink \cite{Musk2019} or the Internet of Brains \cite{Pais-Vieira2015} are a perfect illustration of the potential opportunities ahead.
In fact, the ideas behind these technologies are strongly aligned with the vision of the 
\ac{6g}~\cite{ROUT01}, which is expected within ten years from now.

\begin{figure*}[t!]
\centering
  \includegraphics[width=.7\textwidth]{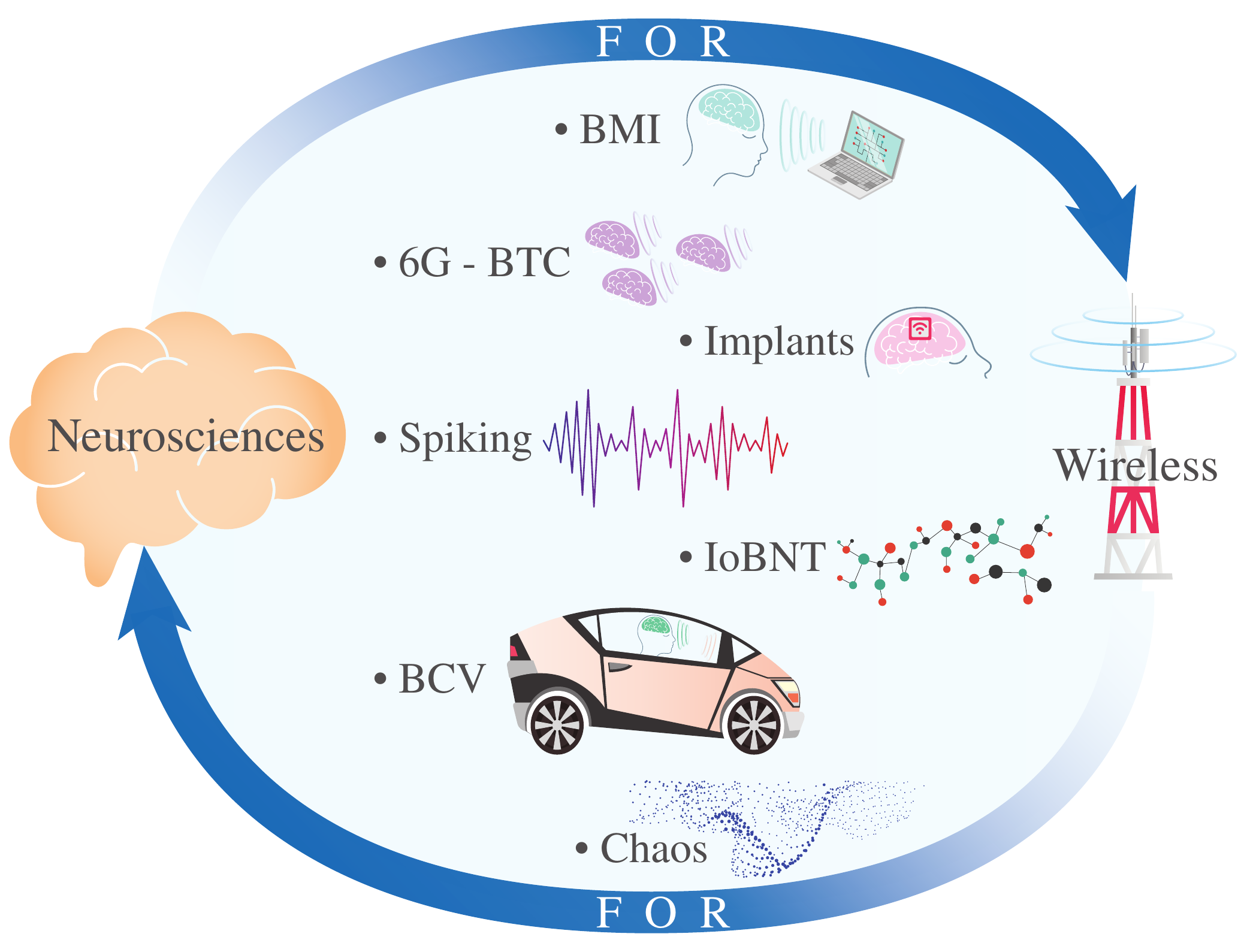}
  \caption{Illustrative picture of the proposed contribution along two threads: Neurosciences for Wireless and Wireless for Neurosciences. The \ac{6g} development should bring \ac{btc} and spiking networks, and also advanced applications like IoBNT and chaos-based communications. A special example is the BCV where brain signals are used to support operation of vehicles.}\label{fig:illust}
\end{figure*}

One of the key drivers of \ac{6g} is wireless brain-machine interactions based on \ac{bmi} enabled by a mobile network designed to support a new type of service, that we call \textit{\ac{btc}}, which can have many contrasts and synergies with the human- and machine-type communications of previous and current generations (4G or 5G).
This approach would allow for more direct interactions between users and networks as compared to current systems, which are dominantly mediated by smartphones.
New services supported by wireless \ac{bmi}, such as interacting with the environment with gestures, motor intentions, or emotion-driven devices, impose remarkably different performance requirements from the current fifth generation systems (5G) in terms of \ac{qope}.
The list of applications is extensive, to cite but a couple of examples: wireless-\ac{bmi}-connected intelligent vehicles, neural-based wireless networks with sensors and actuators working as an ``artificial brain'', as well as the future evolution of virtual reality services~\cite{MINGBIP,CC01,MING01}.

The main contribution of this paper is a novel, holistic tutorial that focuses on this new, promising research field that is located at the frontier of the two established domains: Neurosciences and Wireless Communications.
Our goal here is to provide a tutorial of the state-of-the-art of those fields, mapping the most relevant activities and how they have a great potential to converge with \ac{6g}.
In particular, we delineate the foreseen future applications and their challenges in two threads: \textbf{Neurosciences for Wireless} and \textbf{Wireless for Neurosciences}.

The first one refers to how current and new scientific/technological developments arisen from neurosciences can be employed as part of wireless systems. 
This covers topics from direct wireless brain implants to complexity metrics and spiking solutions for sensor networks.
The second topic refers to how wireless communication technologies (mainly \ac{6g}) and fundamental limits can support neurosciences' research and technological development.
Topics in this thread include how communications/information theory can provide the fundamental limits of neuronal communications, which have chaotic nature.
We also present a case study -- \ac{bcv} -- that we have identified as an illustrative application that would benefit from the proposed merger between \ac{6g} and neurosciences.

Fig. \ref{fig:illust} presents the key ideas and topics covered by this paper, mapping the future relations between wireless \ac{6g} and neurosciences. We envision an interplay between the two topics supporting the development of \ac{btc}, widespread \acp{bmi} integrated with nanotechnology, chaos-based communication, among others.  
%
All in all, we expect that this contribution can pave the way to a fruitful collaboration between researchers active in brain research, complexity sciences, and wireless communications. In addition, it will provide a single reference that symbiotically integrates the rather disparate state-of-the-art contributions in these two fields.

The rest of this paper is organized as follows.
Section \ref{sec:back} provides the required background about neurosciences and brain research, specially discussing on how brain signals are expected to be part of \ac{6g} systems.
Section \ref{sec:NforW} organizes how neurosciences are contributing to \ac{6g} development, providing details and challenges of wireless brain implants, also describing how \ac{isn} based on spiking signal could build an artificial brain.
Section \ref{sec:WforN} presents the potential advantages that \ac{6g} may bring to neurosciences, considering potential new generation of \acp{bmi} based on \ac{6g} and even the \ac{iobnt}, as well as theoretical and practical approaches related to the chaotic nature of neuronal communications.
Section \ref{sec:case} introduces \ac{bcv} as an existing application that would greatly benefit from \ac{6g} and neurosciences synergistic research proposed here.
Section \ref{sec:final} summarizes this paper pointing out our perspective for future research and technological development.

\section{Background}
\label{sec:back}
Evolution has shaped the animal brain to entail individuals with rapid, robust responses to multisensory, possibly conflicting stimuli, thus ensuring survival. We begin this section by highlighting brain design principles, with a focus on properties with direct relevance for wireless systems. We proceed by describing current implant technology for interfacing with the brain, and then we conclude with a review of key concepts from and current stage of \acp{bmi}.

\subsection{Brain design principles}

The brain is a complex organ, notably composed by nerve cells (neurons) but also by supportive cells, such as glia. Ultimately, one may attribute the diverse components, structures, and dynamics found in brains from different species \cite{Hofman2012,Sousa2017} to singular evolutionary pressures. 

Brains vary with animal specie, weight, and size in non-intuitive ways. For example, cows and chimpanzees both have brain mass in the order of 400 g, despite the notable difference in body mass and cognitive abilities. Heavier by only 10 g, a macaque monkey brain has about 4 times more neurons than that from a capybara \cite{Herculano-Houzel2012}. What may thus account for a large part of human intellect is its 86 billion neurons distributed in 1.5 kg of brain mass. 

Brain regions that are mainly made up of electrically insulated neuron axons (myelinated) are referred to as white matter whereas neuronal cell body is found in the gray matter. From a communication systems perspective, white matter may be seen as insulated wires connecting widespread neural populations from the gray matter. The probability of two cortical neurons being connected is 1 in 100 within a vertical column 1 mm in diameter, and 1 in 1000000 for distant neurons; also, forty to sixty percent of brain mass volume is due to wiring (as a comparison, the volume fraction of wiring in a computer microchip may reach up to 90\%), and only one quarter of all energy is spent by white matter \cite{Laughlin2003}. The conclusion is that the brain presents local, densely connected neural populations that are sparsely connected often with small-word properties. 

A direct consequence of such connectivity pattern is a disproportionate increase in white matter (wiring) volume as cortical gray matter increases. In fact, it has been shown that the volume of the white matter and gray matter scales following a power law with an exponent of approximately 4/3 \cite{Zhang2000}. Strikingly, the theoretical prediction of wiring volume that would minimize conduction delays, passive cable attenuation, and connection length whilst maximizing the density of synapses is $3/5$, which resembles real data \cite{Chklovskii2002}. Thus, evolution has optimized neural connectivity.

Evolutionary optimization of neural connectivity is certainly constrained by energy consumption. The human brain accounts for $2\%$ of body mass but requires $20\%$ of the total body energy budget \cite{Herculano-Houzel2012}. Nearly half of brain energy consumption is due to spiking activity, arguably the essential method with which neural populations communicate \cite{Laughlin2003}. Single neurons have a physiological upper limit in firing rate in the range of hundreds of Hz \cite{Wang2016}, leading to a potential bandwidth of a few Terabits/s for the whole brain. This limit, however, is never reached due to energy limitation. Considering the human brain metabolism, the average spike rate can be no larger than 1 spike per second per neuron \cite{Herculano-Houzel2011}. 

The locally dense, globally sparse connectivity scheme constrained by the brain energy budget may reduce the signal-to-noise ratio \cite{Niven2008}. Considering that more reliable neurons would require a non-linear increase in energy cost (due to neuron physiology), one alternative is to average out large numbers of (noisy) neurons. But that, in turn, would possibly lead to redundant neuronal activity, which is not energy-efficient, unless the network is able to reconfigure on the fly, suppressing connections that contribute little to good choices and reinforcing (making more efficient) those that do not. This overly simplified description is known as neural plasticity, the capacity of neural networks to modify its connectivity patterns based on correlated neural activity and behavioral feedback \cite{vonBernhardi2017}. Furthermore, neural variability is influenced by internal and external inputs \cite{Churchland2010} and mounting evidence suggests that prediction of sensory stimuli is a key learning component \cite{Friston2010}. In summary, learning from experiencing the world to optimize behavior is a central mechanism that supports brain design principles under a limited energy budget.

From a functional perspective, neural electrical activity is essentially oscillatory \cite{Buzsaki2006}. Whether an emergent phenomenon resulting directly from brain architecture or an evolved mechanism that shaped neural development, it is widely accepted that neural rhythms and transient synchronization are the basis of neural communication and cognitive processing \cite{Singer1999, Buzsaki2010, Fries2015, Lakatos2019}. For instance, several neural disorders, such as Parkinson's Disease, are linked to disruptions in brain rhythms \cite{Uhlhaas2010, Mathalon2015, Halje2019}. Nevertheless, we still lack a comprehensive description about how neurons and neural networks convey information.

\subsection{Neural interface technology} 
\label{sec:implants}

Brain sensing and stimulation can be performed in many scales ranging from sub-millimeter single unit action potentials to centimeter-scale EEG (\textit{Electroencephalography}) of thousands of cells. The clear distinction here in the ability to be invasive or not. The main advantage of being invasive is the closer interface with brain cells, which lead to less noisy readings. For that, brain implants and recording technologies have been developed for decades now. They are generally composed of six different parts: probe, epoxy fill, acquisition \ac{ic}, circuit board, connectors, and external cable \cite{shepherd2016neurobionics}. 

The interface with the brain tissue is actually solely the probe, where the signal travels from or to the acquisition \ac{ic} that sits in the circuit board through the epoxy fill and connector. The cable connects the whole system to an external system. There is a variety of implantable that range from the largest ones, which was just described, to miniaturized versions, where all the implant components are compressed to a physical device that can be fully implanted into the brain. The smaller devices will have an increased number of challenges regarding their implantation procedure (sometimes through open skull surgery), as well as their functionality, biocompatibility, and longevity \cite{wellman2018materials}. 

Implants functionality are impaired by their implantation position, tissue scarring as well as body foreign reactions that promote the denigration or breakage of the probes. These devices must be, therefore, highly biocompatible, which is mostly driven by the materials that are used to fabricate the probes. Fully-immersed implants are also hermetic sealed with biocompatible materials, however, the probes are located outside the sealed environment and must be treated differently. 

Brain implants are known to be bulky, where the longevity of these devices are a major concern \cite{clement2019brain}. Due to the limited studies in the topic of longevity, the characterization of implants long term usage is poorly defined. However, even with the above-mentioned challenges, implants are more reliable sources for precision in sensing and stimulating brain tissue \cite{maguire2013physical}. The main benefits are that action potential units are purely sensed as opposed to the cumulative noisy signals from hundreds or thousands of cells, and also the ability to inject stimulation current at the precise amount for each cells. These type of miniaturized implants are discuss in \cite{im2016review}. However, fully-immersed implants are available as well for areas between $0.5$cm and $1$mm, which respectively represent thousands to hundreds of neurons. For the first we will manipulate signals called \textit{electrocorticography} (ECoG) and the latter \ac{lfp}. The method to either sense and stimulate neurons across the above-mentioned scales are different but here the discussion is solely focused on the device and the types of signals that they deal with.

Fully-immersed implantable that rely on tethers to connect the device to the an external interrogator inhibits long-term usage and reliability as this connection can be broken easily through movement or patient activity \cite{newton2020spinal}. These tethers have additional challenges including lack of scalability as well as greater body reaction. The number of neuron interface channels is also limited to the number of tethers that a system has. Even though the relationship is not direct, as one tether can have many probes, they are not a good choice when multiple areas of the brain are planned to be interfaced with a single system. On top of that, as the targeted area of study is deep in the brain, this will result in larger tethers that are harder to manage. The possibility of eliminating these tether for wireless-based system has risen the interests of many researchers in the area, as well as the opportunity of merging the areas of \acp{bmi} and wireless communications. 

These new set of systems have to account for the many barriers imposed by the brain in order to have functional implants \cite{maguire2013physical}. These barriers will be explained in more detail later in the paper. The devices need mainly now to not only interface with the neurons but have the capability of converting wireless energy into circuit current. This added complexity is nowadays feasible with the recent advancements in microelectronics and nanotechnology, where energy-converting devices are well understood. Now the idea is to bring that into the implantable, which requires understanding the human body as a communication channel. Since the brain is comprised of multiple different tissue types, and each type poses different interactions with the propagated system, the wireless communication system between implantables and external devices, or derivations of thereof, must precisely chose a desired frequency range that enables the planned application \cite{ferguson2011wireless}. The system must be precisely tailored to the application because the variety of diseases are understood to have different characteristics and peculiarities that require different system functioning as a whole. For example, while brain stimulating devices for epilepsy requires the curbing burst-like event in the brain that requires constant stimulation in random short-term periods, for Parkinson's disease the stimulation is constant at a particular rate at different times. The same goes for sensing applications. 

There are many different challenges that are being focused on the wireless implants. The major attention is actually developing prototypes for either primate models as well as freely-moving animals \cite{agha2013fully,borton2013implantable,yin2014wireless,seo2016wireless}. Since most of works found to that are either theoretical or tested in highly controlled laboratories, it is hard to affirm that we know about all the challenges of these systems from a practical perspective. However, wireless implants must counter for: batteryless devices, data rate requirements, networking, and external control. These challenges do not include biological ethical and safety concerns that must be considered. So far, there are many technological breakthroughs to be made in order to claim a functioning wireless \ac{bmi} that is likely to be used in clinical settings \cite{shepherd2016neurobionics}. This major obstacle needs to be overcome first. However, it will be a truly remarkable engineering achievement where its impact is immeasurable, but certainly an everlasting change to humanity.

\subsection{Neural signals}

The previous section highlighted the great variety of neural recording technology. As one would expect, each method provides signals with distinct properties and a comprehensive description would be beyond the scope of this work. Thus, we now describe three neural signals that compose most of electrophysiological works \cite{Buzsaki2012} and that are central to interface \acp{bmi} to modern communication systems. 

\begin{figure}[t]
  \begin{center}
  \includegraphics[width=1\linewidth]{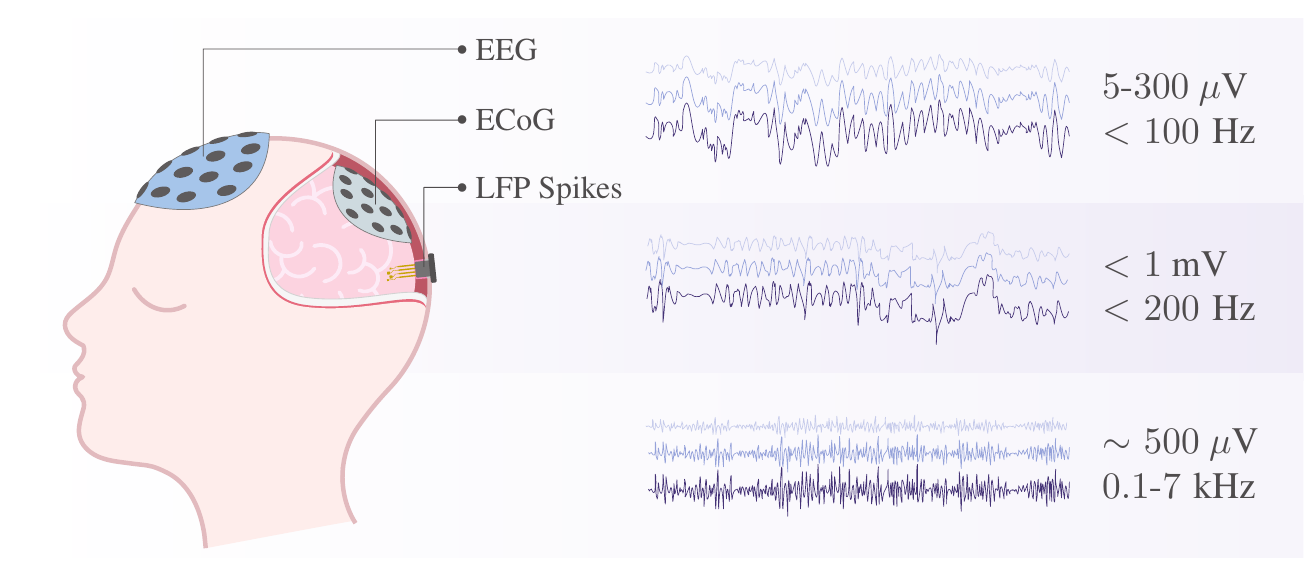}
  \caption{Representation of the most common brain signals used in \ac{bmi}: non-invasive EEG (top right panel), invasive LFP and spike (middle and bottom panels). ECoG signal properties are similar to those from EEG, however, because it is an invasive method, it is far less used in human studies. Figure adapted from \cite{Thakor2012}.}
  \label{figSignal}
  \end{center}
\end{figure}

Considering invasive recording methods, spikes relate to the membrane potential of a single neuron over time. This signal has a strong non-linear dynamics (Fig. \ref{figSignal}) due to neuron physiology and ionic currents flow. Spikes are typically sampled at 40 kHz by multi-electrode arrays, each electrode capturing the resultant membrane potential of surrounding neurons. This multidimensional signal is then fed into a spike-sorting algorithm, responsible for identifying the membrane potential time-series of each individual neuron \cite{Rossant2016}. Next, spike times are identified and saved either as a time-stamp vector (millisecond resolution) or as a binary vector (1 if a spike has occurred, 0 otherwise). The sequence of spikes over time from a single-neuron is known as spike train, which is the data structure used as input to spike-based \acp{bmi} \cite{Lebedev2017}.

The same time-series used to construct spike-trains can be used to extract another signal, the \ac{lfp}. For that, a low-pass filter (<300 Hz) is applied to the raw electrode signal and then downsampled, usually to 1 kHz. \ac{lfp} relate to the superimposing electrical potential of thousands of neurons surrounding the recording electrode \cite{Buzsaki2012}. The spectral power density of this field is inversely proportional to frequency and is transmitted through brain tissue, a phenomenon known as volume conduction. The most common input in \ac{lfp}-based \acp{bmi} are features extracted from \ac{lfp} frequency power spectrum \cite{Aggarwal2013}.

The typical signal used in non-invasive approaches is the EEG. EEG and \ac{lfp} oscillations share similarities \cite{Steriade2003, Buzsaki2012}, but, because recording electrodes are further away from neuronal sources, noise, muscle contraction artifacts, and other tissue-related interference make EEG a less informative signal than invasive recordings. EEG is commonly recorded from 16 to 128 channels, studied at frequencies up to 100 Hz, thus sampling rates rarely exceed 1 kHz.    

\subsection{Brain-machine interfaces}

The rapid progress in neural recording technology (Section \ref{sec:implants}) has paved the way for the development of \acp{bmi} \cite{WOLPAW2002767, Lebedev2017}. A \ac{bmi} is a closed-loop framework, in which neural signals are sampled, pre-processed, and fed into a decoding algorithm (regression or classification) which can map behavioral intents from the brain to artificial devices, whose action outcomes are perceived by the subject sensory systems thereby closing the loop. Applications are diverse, from shedding light into basic neuroscience research \cite{Orsborn2014} to contributing to motor rehabilitation in spinal-cord injured patients \cite{Donati2016}. 

From a technological perspective, \acp{bmi} rely on the continuous progress in electrode design \cite{Hong2019}, data recording \cite{Chen2017}, and signal processing \cite{Shanechi2017}. However, there is a central gap in \ac{bmi} research that is shared by other neuroscience fields: what is the essence of the neural code? In other words, what are the features of neural activity that carry information about sensory stimuli and cognitive behavior? For instance, there is solid evidence for rate and temporal codes \cite{Buzsaki2010,Stanley2013,Lisman2013}, but what exactly are the anatomo-neurophysiological patterns and information processing mechanisms of such codes remain unclear.

Decoding algorithms depend on the signal that it is exploiting. For spike-based \acp{bmi}, most decoding algorithms map spike rate or inter-spike time interval changes into behavioral choices. As single neuron responses vary considerably within and between task trials \cite{Churchland2010}, using recordings from populations of neurons result in more robust interfaces \cite{Nicolelis2009}. If, instead, \ac{lfp} signals are to be used, the common approach is to extract frequency power spectrum features from data blocks over time as the behavioral task unfolds \cite{Aggarwal2013}, given that specific frequency bands have been shown to correlate with behavior \cite{Buzsaki2006}, a fact that also holds for EEG studies.

Finally, neural plasticity, the capacity that neural networks have to dynamically change topological and functional connections based on internal and external stimuli, is fundamental for \acp{bmi} to operate properly \cite{Lebedev2005, Koralek2012, Orsborn2014}. Thus, given that \ac{bmi} design has to carefully consider processing time and sensory feedback delays, \ac{6g} and modern communication systems have plenty to contribute. 

\subsection{\acs{6g} and the brain}

As discussed above, neural signals and their application in \ac{bmi} impose new challenges for communication systems, mainly the established communication systems that are designed to support transmissions related to humans and/or machines, not brains.
In particular, the earliest generations of wireless cellular systems, such as 2G and 3G, sought to connect people via large and bulky mobile phones whose primary function was to deliver voice and short message services. Then, the advent of the smartphone revolution that started with the introduction of the iPhone transformed mobile hand-held devices into powerful and capable computing platforms that can run a plethora of applications ranging from traditional voice services to video streaming, social networking, mobile TV, and mobile gaming. The common denominator among all these applications is that they are used to \emph{connect people} in a variety of ways and, hence, communication among smartphone and similar devices was dubbed as \ac{htc}. However, the past decade ushered in a whole new type of wireless communication dedicated to connecting machines within the so-called \ac{iot} system. Indeed, the emergence of \ac{mtc} links has revolutionized the wireless industry and was the driving force behind the ongoing deployment of 5G wireless systems. At this juncture, it is natural to pose the following question: What type of mobile devices will disrupt the wireless industry and drive beyond 5G wireless systems in the same manner that the iPhone and the \ac{iot} did?

\begin{figure}[t]
  \begin{center}
  \includegraphics[width=1\linewidth]{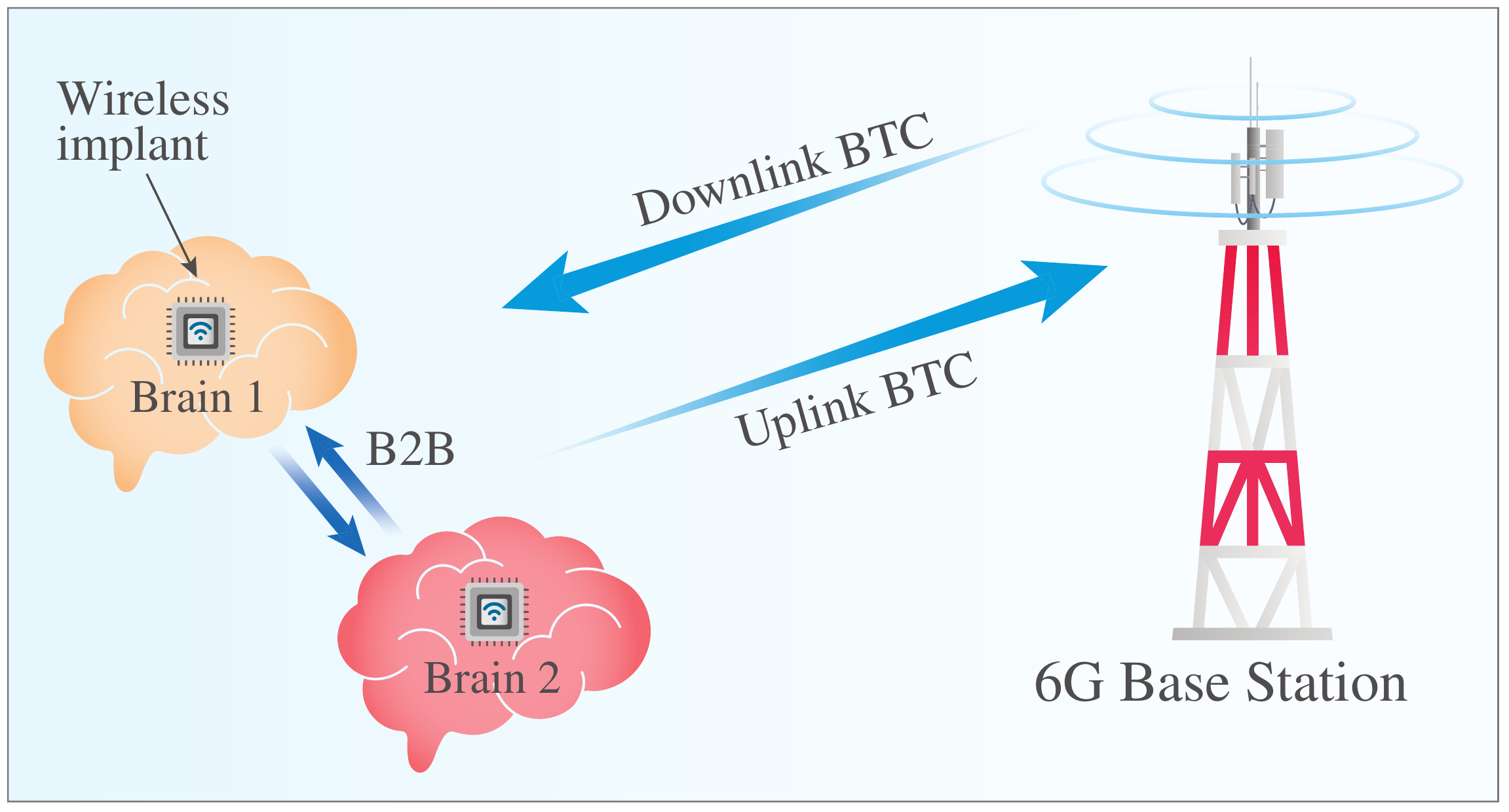}
  \caption{Illustration of different types of \acs{btc} services in \acs{6g} systems.}\label{fig2}
  \end{center}
\end{figure}

Although a conclusive answer to this question is not possible at this time, it is very natural to posit that next-generation wireless devices will no longer be handheld smartphones or \ac{iot} sensors in the field, but they will rather be wearable devices along a human body as well as human brain implants. This observation is not a mere speculation, but it is instead motivated by the tremendous advances that we are witnessing in the area of wearable and human-embedded devices, with Neuralink's recent achievements being a prime example. In addition, the shift toward implants is further motivated by several emerging wireless services, such as immersive \ac{xr} and \ac{bmi}, in which the human body and brain become an integral part of the wireless service~\cite{ROUT01}. In these services, it will soon become necessary to provide communication links among, not only machines (\ac{mtc}) and human users (\ac{htc}), but also among the brains of different users. Hence, we foresee that \ac{btc} will be the next frontier in wireless connectivity, as indicated in Fig. \ref{fig2}. \ac{btc} links must be designed in a way to seamlessly connect a human brain to a wireless network and potentially provide two-way communication among the user's brain implants and the various network and \ac{iot} devices. A unique feature of \ac{btc} links is that they will require the network to match the capabilities of the human brain -- arguably the most powerful computer in the world. Clearly, delivering \ac{btc} will bring forth a whole new set of challenging wireless networking problems that can only be addressed by bringing together tools from neuroscience and communication theory, as well as adjunct areas. Indeed, as articulated in the next section, the use of neurosciences to model and capture the brain's inherent features can soon become an integral component of wireless networks that cannot be ignored when modeling, analyzing, and optimizing the wireless networks of the future.

\section{Neurosciences for Wireless}
\label{sec:NforW}
In this section, we will discuss how the technological developments based on the state-of-the-art in neurosciences will open many new possibilities with related challenges in wireless communications beyond 5G systems.
This will include the support of \ac{btc} and intelligent (neuromorphic) sensor networks based on spikes.

\subsection{Direct brain implants that communicate wirelessly}
Communications with brain implants will be a hallmark of next-generation wireless networks and, hence, we must have a deeper understanding on how to deliver wireless services to networks with \emph{brain-in-the-loop}. To do so, we will first discuss some use cases that highlight different ways in which \ac{btc} will be integrated in wireless networks. Then, we delve into the various challenges associated with the identified use cases and we conclude with discussions of open research problems and some preliminary results.

\subsubsection{Use Cases} The first step toward understanding the unique wireless challenges of \ac{btc} links consists of delineating possible \ac{btc} use cases in an actual network. In this context, we envision three key use cases (as illustrated in Figure \ref{fig2}):
\begin{itemize}
    \item  \textbf{\emph{Downlink BTC:}} \ac{btc} links can be used in the downlink of a wireless network. Here, the downlink transmission links are used to transmit data from the network towards brain implants. A chief use case in this context is \ac{xr} services. Indeed, next-generation \ac{xr} services may tap directly into the human brain cognition in order to provide a truly immersive virtual world that a wireless user can navigate using its brain along with various body-implanted sensors. In such use cases, the brain is the receiver of the wireless data and it must process the received data in order to create the sought-after user experience. It is natural to think that, for applications such as \ac{xr}, downlink \ac{btc} traffic will require high data rates.

    \item  \textbf{\emph{Uplink BTC:} }\ac{btc} links can be used for uplink communications in order to transmit information extracted from the human brain through its implants to other network devices and servers. Uplink \ac{btc} will be particularly important for \ac{bmi} services in which data from the human brain must be transmitted to other devices for various control purposes. Two key \ac{bmi} examples that require uplink \ac{btc} include multi-brain-controlled cinema~\cite{TSM01} in which humans participate in real time in a movie through brain input and wireless cognition~\cite{IOT5G01} in which a drone or autonomous vehicle is controlled by a brain. Naturally, for critical applications that require uplink \ac{btc}, reliable uplink connectivity is necessary.

    \item  \textbf{\emph{Brain-to-Brain (B2B) Communications:}} \ac{btc} links can be used to establish direct communications among the brain implants of different users within the same or different environments. \ac{b2b} communications can be seen as the next step in \ac{d2d} communication whereby now the devices are direct brain implants. \ac{b2b} \ac{btc} links can be useful in many scenarios such as immersive gaming in which players can coordinate via \ac{b2b} links and education in which \ac{b2b} links can be used in teaching classes or conducting projects.

\end{itemize}

\subsubsection{Challenges} Having laid out the key uses cases for \ac{btc}, our next step is to identify the unique challenges of these use cases, compared with traditional \acs{htc} and \ac{mtc} services. First, it is well-known that the bottleneck of \acs{htc} services is downlink communication and the bottleneck of \ac{mtc} services is uplink communication. In contrast, in \ac{btc}, we can easily see that both uplink and downlink may constitute a bottleneck for data rates. On the one hand, to provide immersive experiences, significant data must be downloaded in the downlink towards the brain implants. Meanwhile, in order to provide sensory and control inputs from the human brain to the network and its services, brain data must be transmitted from the implant to the network. At first glance, one would think that the uplink input will still be short packet, small data, as is the case for \ac{mtc}. However, anecdotal results in \cite{IOT5G01} show that the amount of data generated by a brain for wireless cognition services can be in order of terabytes. Hence, uplink \ac{btc} will also require ultra high speeds from the wireless links, which is in sharp contrast with \ac{mtc} services.

Second, despite its immense computational abilities, the human brain has its own perceptual and cognitive limitations. These cognitive limitations can be affected by multiple human brain sources such as context, attention, human fatigue, or limited cognitive abilities. From a wireless perspective, these cognitive limitations can be translated into limitations on the way in which a human brain perceive network \ac{qos} metrics such as rate or delay. For example, as shown in \cite{NEURO00}, due to its architecture and neural network dynamics, the brain may exhibit intrinsic time delays that affect the way in which it perceives the world around it. Therefore, a key challenge here is to develop new techniques from  neuroscience in order to provide new models for the brain that can quantify these limitations and potentially be used in a wireless network framework to map those limitations into \ac{qos} or \ac{qoe} metrics. Note that here this challenge significantly differs from traditional \ac{qoe} metrics such as the mean opinion in which one can simply use interviews or basic experiments to quantify \ac{qoe}. Instead, here we need to quantify the so-called \ac{qope} introduced in \cite{ROUT01} in which the specifics of a human's physiological characteristics, particularly the brain, must be captured and mapped into the conventional wireless QoS metrics. 

Third, 5G systems are expected to deliver three broad types of services: \ac{embb} services in which high data rates are expected, \ac{urllc} services in which reliable low latency transmissions are required for services such as \ac{iot} sensing that do not require high rates, and \ac{mmtc} that deals with the connectivity of a massive number of \ac{iot} devices. Traditionally, these service classes are expected to be distinct for one another. For example, \ac{urllc} services are assumed to not require any data rate guarantees since they deal with short-packet transmissions of \ac{iot} sensor data. Meanwhile, \ac{embb} services simply require high rate and do not need much reliability or low latency guarantees. In contrast to these traditional service classes, \ac{btc} services may require, simultaneously, high reliability, low latency, and high (\ac{embb}-level) rates. Wireless cognition provides an example, which remotely controlling an autonomous vehicle via the brain will necessitate very high reliability and very low latency, due to the criticality of the circulating data. Meanwhile, this remote control will also require very high rates as discussed in \cite{IOT5G01}. Hence, when dealing with some \ac{btc} services, it is necessary to provide both \ac{embb}-level rates and \ac{urllc} reliability and latency, which is yet another key challenge. Moreover, as the technology becomes more mainstream, we can anticipate a massive numbers of \ac{btc} links active at a given time and, hence, in this case, \ac{mmtc} features will also appear, particularly for \ac{b2b} links. Clearly, the evolution towards \ac{btc} may require us to revisit the existing 5G distinction among different services.

Fourth, although \ac{b2b} \ac{btc} links share many of the aforementioned challenges, they also bring a new dimension which has to do with the interactions among human brain, which have different physiology and cognitive capabilities. Addressing this challenge requires a better understanding of networks of brains and how they may interact with one another. Naturally, \ac{b2b} communications brings in a suite of interdisciplinary challenges that require a better understanding of not only the communication features of \ac{b2b}, but also the potential interactions among the brains of different users whose context, demographics, and characteristics are disparate.
A comparison between \acs{htc}, \ac{mtc} and \ac{btc} is presented in Table \ref{table: htc_mtc_btc}.

\begin{table*}[t!]
	\centering
	\caption{MTC versus \acs{htc} versus \acs{btc} features and requirements in the context of \ac{6g}}	
	\begin{tabular}{p{3cm}p{3cm}p{3cm}p{3cm}} 
		%
		\textbf{Requirements/features} & \textbf{HTC over cellular} &
		\textbf{MTC over cellular} &
		\textbf{BTC over cellular}
		\\ \hline\hline
	\textbf{	Uplink} &  
		High signaling overhead to overcome features such as mobility (handover).  Relaxed/mild throughput requirements. &
		Reduce signaling overhead (e.g., via fast uplink grant \cite{SAMAD2}) and specialized random access procedures to support high users density.  Characterized to consider short package - small data. &
		Brain control \ac{btc} may require reliable and low latency transmission with varying throughput (low to high). Some special cases have high throughput requirements with high signaling overhead to support mobility.\\
		\hline
	\textbf{	Downlink} &
High throughput requirement for individual users &
Most typical applications demand low throughput for individual users which is mostly used for software update/upgrade of the embedded systems.&
High throughput requirements with high signaling overhead to support mobility. \\
		\hline
	\textbf{	Subscriber load} &
Few users that are mostly supported by small cells &
High density  of devices &
High user density is expected with URLLC and high data rates simultaneously.\\
\hline
    \textbf{    Device Types} &
Broadband devices as smart-phones, tablets, personal computers.&
Devices are mostly small (IoT) devices carrying sensors with specific power-constraints.&
Brain implants (invasive and non-invasive).\\
\hline
\textbf{        Delay requirements}&
It varies between best-effort services as e-mails, and high-definition games to support human sensitivity in terms of latency. &
Many  delay-tolerant applications and also low latency ones, this last mostly employed in closed-loop control systems, protection systems, and other critical application. &
Strict delay requirements for both, downlink and uplink. \\
\hline
        \textbf{Energy requirements}&
Relatively high energy consumption since it supports many applications that run on a smart phone. In most of the cases the battery is charged once per day in average.&
\ac{mtc} devices usually perform specific tasks that provide flexibility to setup the data transmission duty-cycle. In many cases, the battery should be re-charged/replaced long periods of time as 5 years.&
High energy efficiency is needed to support a high duty cycle demand such as, for instance, the case of \ac{b2b} communication.\\
\hline
        \textbf{Signaling requirements}&
Accurate signaling protocols are required due to uncertainties related to mobility and data usability. &
Reduced requirements in UL in order to support massive number of users.&
High level of signaling for both uplink and downlink is required  to support URLLC, high data rates, and massive users simultaneously.\\
\hline
	\end{tabular}
	\label{table: htc_mtc_btc}
\end{table*}

\subsubsection{Research Problems} Clearly, the aforementioned challenges bring forward interesting research problems at the intersection of neuroscience and wireless networks. In general, providing wireless networking with ``brain-in-the-loop'' is a rich research area with many open problems that follow directly from the identified challenges.

One of the first open problems in this area pertains to the need for new techniques that combine neuroscience with wireless network modeling in order to precisely quantify \ac{qope} measures. On the one hand, one can take a data-driven approach to this problem and look for new machine learning techniques that can dynamically build \ac{qope} metrics by learning from the network's users and their brain behavior. Naturally, the primary limitation of this approach is that it will require significant datasets and long-term observation. However, as datasets in both the neuroscience and wireless communities, are becoming more accessible, we anticipate new opportunities for designing \acp{qope}. On the other hand, one can forego the data-driven approach or complement it with an analytically rigorous approach to model the brain's features. In particular, one can leverage existing tools from control theory and neuroscience to view the brain as a control system with a feedback loop and, then, use this observation to quantify how different input (from the wireless network) are translated into meaningful information for the brain. We can potentially study the transfer function of this brain control system and understand its behavior with respect to different input excitations coming from a wireless network. Using this approach, we can potentially investigate how \ac{qos} metrics are translated into \ac{qope}. This can benefit from some of the existing studies on how to look at the brain's control signals (e.g., see review in \cite{MPM}). Last, but not least, real-world experiments with actual participants can be organized to better understand how the brain perceives \ac{qos}. These behavioral experiments can be combined with behavioral frameworks, such as prospect theory and cognitive hierarchy theory~\cite{PT00,WSM01,PT02,CHT00}, that quantify how humans make decisions to yield new insights on how to model the response of a brain to wireless signal inputs, for different services.

Moreover, as discussed earlier, there is a need to calculate the processing power of the brain, using neuroscience techniques, so as to truly quantify the amount of data needed. Here, instead of looking at brain limitations, we are more interested in the brain \emph{capabilities} and how the capabilities can impact wireless communication. While the back-of-the-envelope calculation of  \cite{IOT5G01} provides a first step in this direction, there is a need for more rigorous modeling that takes into account realistic brain models or real-world brain data. 

Once \ac{qope} metrics are developed and brain capabilities are quantified, a very natural next step is to investigate how network management, multiple access, and network optimization techniques will change when dealing with \ac{btc} links and \ac{qope}. In particular, one can design new \emph{brain-aware} resource management techniques that can tailor the network resources and operation to match the brain's required performance while also being cognizant of the brain's capabilities as well as its inherent limitations in processing information, in general, and processing wireless \ac{qos} metrics, in particular. One fundamental question that we can pose in this area pertains to whether or not brain constraints lead to a ``waste'' of wireless resources due to the delivery of a \ac{qos} metric that cannot be perceived by the brain. For example, it is natural to ask whether a human brain can see a difference between two different delay values, i.e., will $10$~ms be perceived as a better \ac{qos} than $20$~ms? 

Moreover, the co-existence of \ac{btc}, \acs{htc}, and \ac{mtc} links, which is expected in early-on deployments of beyond 5G cellular systems will bring forth a rich set of resource management questions pertaining to how one can enable a seamless co-existence of these fundamentally different service classes. Here, beyond investigating radio resource management problems, we can also investigate new ways to incorporate brain features into network slicing problems. Indeed, network slicing must now handle a new type of services and hence a rich set of new open problems can be observed. Moreover, since \ac{btc} links carry characteristics from all three traditional 5G services, i.e., \ac{embb}, \ac{urllc}, and \ac{mmtc}, it is necessary to investigate how one can guarantee high rate, low latency, and high reliability simultaneously, in presence of a potentially large number of \ac{btc} links. Here, one can start by first identifying the achievable performance of \ac{btc} links over 5G and beyond systems (e.g., over terahertz or millimeter wave systems). In particular, there is a need to analyze the rate-reliability-latency operation regime that can come out of the deployment of \ac{btc} links over a cellular system and then to translate this analysis into a feasible \ac{qope} regime of operation that maps the rate-reliability-latency requirements into \ac{qope} measures. Once this feasible \ac{qope} regime is identified, one can revisit traditional problems of multiple access in order to see how all three factors: reliability, latency, and rate, can be matched to the requirement of both the user's brain and the network service that is being adopted. Indeed, here one  important direction is to study how different types of services (e.g., \ac{xr}, \ac{bmi}) will have different brain and \ac{qope} requirements.

Another important open problem is to quantify and measure information from brain implants. Here, information is no longer standard digital information, but instead, it is now directly the byproduct of a user's brain and, thus, its characteristics will be different than standard information. Hence, using tools from fields such as information theory, we must study how ``information'' can be modeled when its the output of a brain (e.g., using information-theoretic perspectives). Then, we can revisit the recently introduced concepts of \ac{aoi}~\cite{Yates17,BO00,BO01} and \ac{voi} and see how these metrics change when dealing with a brain network. For example, we can observe that the way in which information ``ages'' when it is transmitted among brains may no longer be linear, as is the case for traditional wireless information transmission. In this respect, aging of brain information transmitted over \ac{btc} or \ac{b2b} links may require new approaches that depart from the classical linear aging process that is used in most of the \ac{aoi} literature. Here, it is necessary to investigate how information propagate in a brain (e.g., using models such as those in \cite{NEURO00}) to see how timing delays and the neural composition of the brain capture and process information. Similarly, new ways to quantify \ac{voi} when it is the output of a brain are needed. Once information is quantified and its different metrics are revisited, we can leverage this analysis for both physical layer designs as well as for routing and information flow problems, as discussed next.

At the physical layer, the deployment of \ac{btc} will require new designs at the wireless physical layer. For instance, it is interesting to investigate whether new waveforms can be designed so as to translate the brain output into meaningful signals that are aware of the unique features of the brain. Here, we anticipate a need for merging tools from neuroscience, information theory, and communication theory. In some sense, we must investigate how the brain information that is quantified using information theory techniques can be translated into digital waveforms. For example, here, by taking the control-theoretic approach for modeling the brain, we must investigate how the output of the control system model of the brain can be translated into a digital communication signal that can be transmitted over \ac{btc} links.

Finally, in terms of routing and information flow, it is necessary to develop new techniques to manage \ac{b2b} and \ac{btc} links in a way to optimize \ac{aoi} and \ac{voi} metrics when those metrics pertain to a brain. As already discussed, the models for these two metrics will be significantly different when dealing with human brains. As such, existing latency-optimal or rate-optimal routing and information flow algorithms will not necessarily be \ac{aoi}-optimal or \ac{voi}-optimal. Hence, we envision many fundamental routing problems that can now see a wireless network as an overlay of two inter-related systems: a) a human-to-human network that receives and translates information through a brain and b) a \ac{d2d} network that carries this information. Modeling the relationship between these two systems and integrating it into network routing and information flow optimization problems is clearly an important and meaningful open-problem that brings together neuroscience, communication theory, and network science.

\subsubsection{Sample Results} The area of wireless network design for incorporating \acp{btc} is still at its infancy and hence not many works have looked at related problems. However, in \cite{HLOOP00}, we have made a first step in this direction by analyzing how to use a data-driven approach that uses user brain information to create \ac{qope} measures that map wireless delays into brain perceptions and, then, those perceptions are integrated into a resource allocation problem. In this early work~\cite{HLOOP00}, to model \ac{qope}, we explored the observation in \cite{NEURO00} that the brain can have multiple ``modes'' depending on the age, sex, demographic, time of day, and other social features, to learn how the brain perceives delay in a wireless network. The \ac{qope} therein pertains to how one can translate a brain mode (extracted from the data) into a perception of \ac{qos} metrics such as delay. Our work in \cite{HLOOP00} showed that, for a wireless network, the aforementioned brain mode limitations makes the wireless user unable to distinguish the \ac{qos} differences between different wireless delays. In other words, the \ac{qope} of a user maps each delay value to a different brain perception. Building on this observation, our results in~\cite{HLOOP00} show that due to the cognitive limitations of the brain, delivering ultra low latency for services such as \ac{xr} may not improve the user experience, because, at a very low latency regime, the user's brain can no longer distinguish the difference between different delays. For instance, our results show that it is less probable that a user distinguishes between 20 ms and 10 ms delay compared to 30 milliseconds and 20 ms. As such, when designing \ac{btc} links, one key challenge is to properly model and capture the limitations of the brain and factor in those limitations into the wireless network design.
\begin{figure}[!t]
  \begin{center}
   \vspace{-0.2cm}
    \includegraphics[width=\columnwidth]{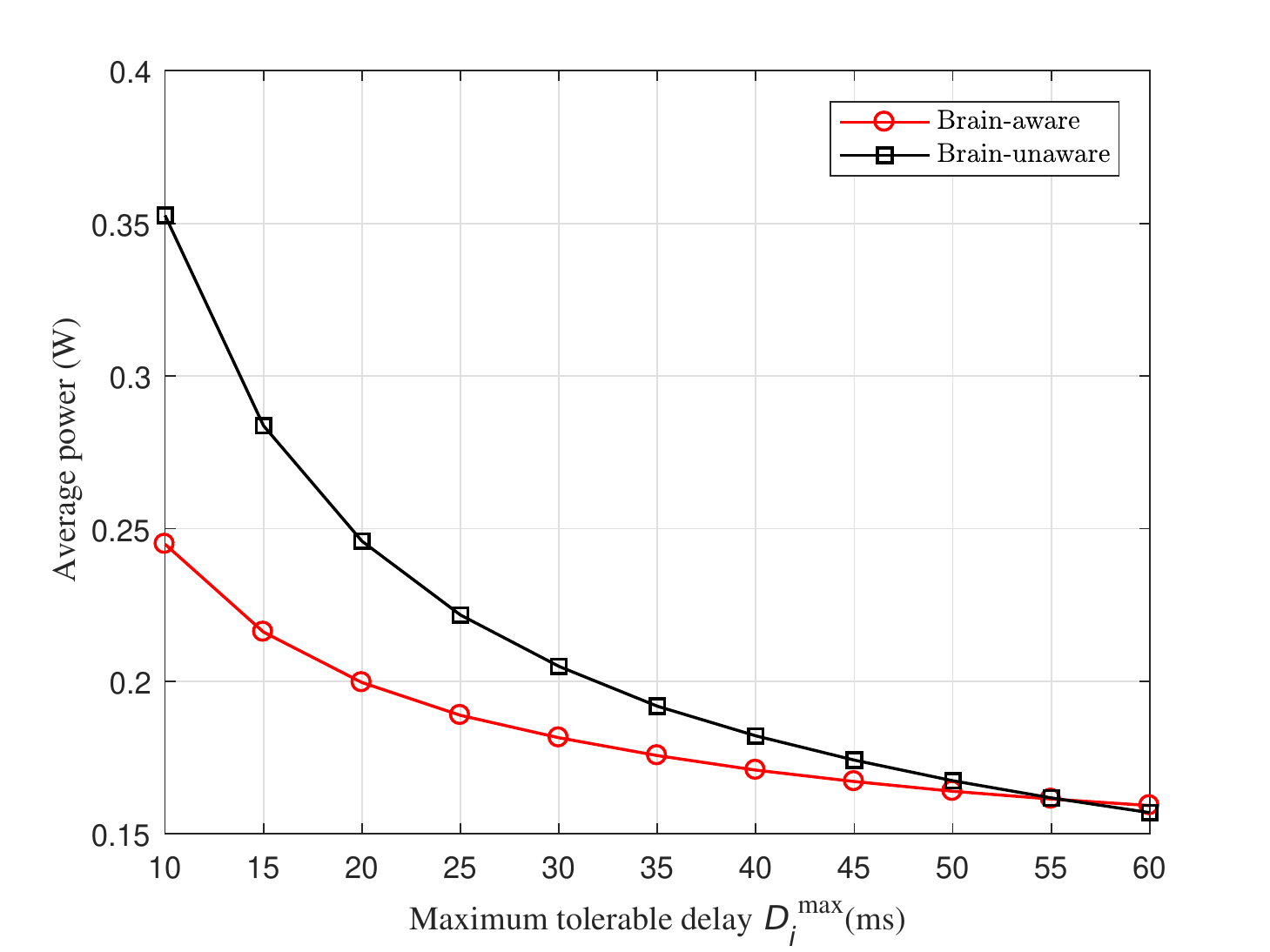}
    \vspace{-0.3cm}
    \caption{\label{fig:brain} Early result from~\cite{HLOOP00} showing how a brain-aware resource management approach can save significant resources (in terms of power), particularly at a low latency regime, by being aware of the cognitive limitations of a brain that limits its perception of delay. The x-axis here represents the ``raw'' maximum tolerable delay threshold by each user.}
    \vspace{-0.2cm}
  \end{center}
\end{figure}

In addition, in~\cite{HLOOP00}, we have then incorporated the learned brain limitations into a downlink power control problem with brain perception constraints. We did so in order to test the hypothesis that brain-aware resource allocation approach can significantly save network resources. Here, we have particularly shown that, by explicitly accounting for the cognitive limitations of a user's human brain, the network can better distribute resources to \ac{btc} users that need it, when they can actually use it. This is in stark contrast to conventional brain-agnostic network resource allocation techniques in which resources may be wasted, as they are allocated only based on application \ac{qos} without being aware on whether the human user's brain can realistically process the actual service's raw \ac{qos} target. For instance, in Fig.~\ref{fig:brain}, extracted from our work in~\cite{HLOOP00}, we compare the performance, in terms of power allocated to optimize the wireless system while meeting delay threshold and reliability constraints, between a brain-aware resource allocation approach and  a brain-unaware resource allocation approach. Strikingly, this figure shows that at very low latencies (below $40$~ms), a brain-aware approach can save significant resources by being aware that the brain of a user (depending on the mode of the user) may not distinguish a \ac{qos} difference between different values of latency. Clearly, these promising results can be used as a building block for new research in this area that can potentially address the rich set of open problems previously identified.

\subsection{Brain barriers for wireless channels}
The physical medium also imposes challenges to any wireless system that would support \ac{btc}.
Remarkably the transcranial wireless channel presents many challenges to the many wireless technology options due to the its structure and function. The brain is covered by the skull and surrounding head tissue that absorb or scatter high frequency signals. 

Lower frequency signals are known to cause either implantation of large devices and head heat increase. Novel wireless solutions most cope smartly with those unwanted effect, but we must take into account that single neurons are known to have high data rate demands for sensing purposes. Here we explore the previously listed brain barriers in \cite{maguire2013physical} focusing our discussion on the wireless technologies as well as the communication channel between implants and external devices.

\subsubsection{Spatial-temporal resolution} The number of neurons and other brain cell types goes beyond the billion unit mark and is considered the biggest challenge in measuring the whole brain information with existing technology and infrastructure. Naive estimates of the whole brain recording lower bound data rate is about 100 Gbits/s, which is already a challenge for today's wireless technologies, let alone for future \acp{bmi}. The forthcoming technologies must include compression techniques that minimize the transmission burden of single action potentials. The compression technique will have an interesting interplay with the sampling rate of signal recording as well as the wireless technologies and their equivalent data rates.

\subsubsection{Energy dissipation} The propagation of transcranial wireless signals that are transduced by implantable devices will result in energy dissipated through the tissue. This energy will be converted to heat, which is also dissipated. Due to the brain's tightly packed structure, damage can occur due to a minimum temperature increase of above two degrees Celcius. Wireless signals, however, can be easily modulated in order to operate bellow 100\% duty cycle of the system's operation, which can help prevent damaging energy dissipation. The brain also present natural cooling mechanisms that can help restore brain normal temperature. However, the real challenge lies in the large scale deployment of heavy and dense recording and stimulation techniques for high-spatial resolutions. Unfortunately, more research needs to be done in order to provide maximum control of energy dissipation.

\subsubsection{Volume displacement} The insertion of devices in the Brain can cause its volume to increase leading to damage of its functioning tissue. A displacement bigger than 1\% of the total brain volume is not allowed for implantables. Wireless technologies can help these implantables to remain in very small sizes either using high frequency transmission or low frequencies for applications in the cortex. The underlying issue is that most effects that inhibit the implantable well functioning happen way over its implantation phase, which the most well-known is the Glial scarring\footnote{Glial scarring is the formation of Glial tissue around the implant preventing its interface with neurons.}. Wireless interfaces can help long-term implantation in this case by allowing that these devices are package within biocompatible material that prevent body foreign reaction to happen, such as the above-mentioned. Future techniques such as \ac{mimo} wireless systems for implantables might help the usage of low frequency solutions for deep brain interfacing that is essential for integrating existing wireless system platforms to future wireless brain interfaces.

\subsection{Brain implants assisted by intelligent reflecting surfaces}

As discussed before, despite the increased risk of injuries and other related issues, invasive wireless brain implants exhibit numerous benefits in comparison with conventional over-the-scalp solutions. It has been shown that these prosthetic devices are capable of sensing more accurate brain activity, interacting directly with the brain, and providing a higher \ac{snr} \cite{ArtRef01}. These capabilities make them powerful tools for enabling \ac{bmi} in future \ac{6g} and beyond. However, before this technology becomes available to the global population, many limiting issues need first to be addressed.
Following the points raised in the previous subsection, one important impairment of wireless brain implants is related to the strong signal attenuation due to tissue blockage and absorption. The high quantity of water molecules in the human body can interact with the electromagnetic waves, absorb a significant part of transmitted power, and distort the radiation pattern \cite{ArtRef02}. Such a  characteristic can deteriorate the communication link and impact reliability. One could, to some extent, alleviate this issue by allocating a higher transmit power; however, this parameter cannot be increased indiscriminately. First, there are strong power restrictions due to human health, and second, in general, the implanted devices have limited access to energy resources. Therefore, new energy-efficient strategies for improving wireless transmission performance in brain applications are required.

In particular, \ac{irs} have recently arisen as appealing devices for smartly controlling the electromagnetic propagation environment. An \ac{irs} consists of a two-dimensional structure that comprises a large number of nearly passive sub-wavelength metamaterials elements with tunable electromagnetic properties. These elements can be dynamically configured to collectively change the behavior of impinging wavefronts so that capabilities like steering, polarization, filtering, and collimation can be achieved \cite{ArtRef03}. Such features make the \ac{irs} technology attractive for improving the performance of wireless communication in brain implants. Conceptually, if the prosthetic implanted device is assisted by an \ac{irs}, it could send and receive information more reliably without increasing its power consumption. This improved brain communication system could be implemented, for example, by implanting \acp{irs} between the skull bone and the skin scalp for assisting sensors and actuators implanted deeper in the brain. In this architecture, the deeper implants would exchange information directly with the brain, while the \acp{irs} would assist the wireless communication established with external devices. An \ac{irs} assisted \ac{bmi} can become able to provide the following capabilities:
\begin{itemize}
    \item \textbf{Improved reliability in wireless data transfer:} by proper tuning the \ac{irs}'s elements, the signal transmitted from the brain implants can be boosted so that a higher \ac{snr} can be achieved at an external receiver, thereby, improving the communication reliability.
    \item \textbf{Reduced power consumption in the brain implants:} since the \ac{snr} can be improved with the help of \acp{irs}, one can decrease the transmit power at the brain implant and still achieve a satisfactory communication performance. This would reduce energy supply requirements and prolong the battery life of the implanted devices.
    \item \textbf{Improved communication security:} since brain implants can both sense and stimulate the brain, security issues become a critical concern in \ac{bmi}. An \ac{irs} can also be beneficial in this context, it can null out information leakage at a potential eavesdropper, or it can operate in shield mode to avoid brain hacking; that is, by properly optimizing the \ac{irs} elements, transmissions to the brain implants coming from a hacker can be completely absorbed/blocked.
\end{itemize}

All in all, the development of \ac{irs} technologies in the years to come combined with brain implants would support the development of \ac{btc} applications in \ac{6g}.

\subsection{New generation of sensor networks}
Another interesting line of research that would be relevant to \ac{6g} systems is brain-inspired (neuromorphic) solutions based on distributed low power sensors that would individually work like neurons but together build an intelligent system (like an artificial brain) \cite{young2019review}.

\subsubsection{Intelligent sensor networks}
Early sensor networks were built using dedicated sensors and fixed network connections to a central processing unit, severely limiting their deployment options. With the addition of wireless interfaces, the area of \ac{wsn} has seen an explosion in possible applications and variety of technologies, and in particular ultra-low power wireless sensor networks – with sensor nodes running for a long period of time on small batteries or even using energy harvesting, have opened the way to drastically expand the capability of connecting the world and everyday things to the Internet. A vast amount of research has been executed on energy management in \ac{wsn}, addressing both energy provision (be it battery driven or using energy harvesting techniques) and energy consumption \cite{Khan2015}. As the data transmission and reception parts of these wireless sensor nodes typically consume most part of the energy, most research has focused on new radio technologies, designs for duty-cycling, and link and networking protocols. However, some research has considered reducing the data that needs to be reported by using prediction-based monitoring or model-driven data acquisition \cite{Dias2016}. Time-series forecasting using moving average or auto-regressive moving average methods are simple and lightweight to implement, and can provide satisfactory results for simple measurements. In more complex cases of high bit rate, heterogeneous, and diverse sensors on a single node, there is a growing need to include more sophisticated data processing on the sensor nodes, to try to limit the amount of data that needs to be transmitted over the wireless interface. Research on kernels for specific data transformations (e.g., \ac{fft}, matrix multiplication) has led to several types of microprocessor optimizations for the \ac{iot} \cite{Adegbija2018}. However, all these approaches are data-focused: the main idea is to get the data from the sensors to a central entity for further processing and analysis. 

For a next phase of truly \ac{isn}, the sensor nodes should contain information-focused techniques that process and convert the data from the sensors into higher-level information elements, before transmitting over the wireless interfaces. If the processing can be done in a power efficient way, the amount of data that needs to be transmitted over the wireless link is only a fraction of the sensor data rate. For example, if a \ac{bmi} application needs to distinguish specific patterns in the EEG signal of a person, instead of compressing the data and sending it to a more central point for analysis, an alternative approach would be to either (i) extract the meaningful features from the data on the sensor node itself, and only send the feature values upstream over the wireless interface, or (ii) do the full recognition on the sensor and only send the identification of the pattern over the air interface.

\subsubsection{Spiking neural networks}

The domain of data-driven learning of models that can perform end-to-end feature learning and classification has been dominated by deep \ac{ann} over the past decade. However, the main focus has been on high accuracy, and typical models contain many neurons (tens of millions), require large amount training examples (tens of thousands), and consume an enormous amount of power (hundreds of Watts) both for training and inference \cite{Li2016}. As the brain contains many more neurons, but still only uses 20W maximum, researchers have turned to investigating new brain-inspired ways of implementing these \acp{ann}. 

A very promising approach are \ac{snn} (also called the 3rd-generation of \acp{ann}) in which communication between different artificial neurons is – like in the brain – performed by passing spikes. Main advantages with respect to energy consumption are (i) spikes use very little energy, (ii) processing only happens when needed, and (iii) spike processing can be done at very low latency. Initial investigations have resulted in dedicated \ac{snn} processors for network sizes up to 100k-1M neurons at below 100 mW power consumption \cite{Merolla2014}, \cite{Davies2018}. While these results already are a huge improvement over standard deep learning architectures, several challenges remain for this technology as described next.

\begin{itemize}

    \item \textbf{{Ultra low power inference:}}
Depending on the application, power consumption for next generation \ac{snn} inferencing should still be reduced with one or two orders in magnitude. This to either run small to average sized networks at extreme low power (e.g. below 1 mW), or build bigger and more powerful \acp{snn} within a slightly higher power budget. Several promising directions are being explored, including: (i) new \acp{ann} to \ac{snn} conversion schemes, using only a minimal amount of spikes in temporal coding schemes, instead of the initially used rate coding schemes \cite{stckl2020classifying}, (ii) ultra-low power wake-up nets that can do an initial detection of interesting events at very low power, and possible wake-up a bigger \ac{snn} to analyze the event when needed (this idea is analogous to ultra low power wake-up radio solutions that enable a sensor node to listen to incoming data a very low power, and only wake up the full received when needed \cite{Piyare2017}); (iii) new materials for compute beyond current CMOS technology \cite{Nguyen2017}. New insights from neuroscience in different neural coding schemes and network architectures can be an inspiration for new highly efficient, low latency, and ultra low power solutions.

    \item  \textbf{{On-device learning:} }
While on-device inference can be very powerful for detection generic features, it can suffer from being a static model, most often trained in the cloud on aggregate data sets, and subsequently deployed on the sensor nodes. Many applications either need to function a changing environment, or need to adapt to the characteristics of an individual user (e.g. \ac{bmi}). In this case, learning of characteristic features should be possible on the sensor device itself, from limited data, and in real-time. Deep learning networks are not very suitable for this as their learning process typically requires huge amount of data samples, training takes a very long time, and is very computational intensive. New techniques based on learning mechanisms of the brain (e.g., \ac{stdp} variants or \ac{htm}) have already lead to promising results with local learning rules than can function on-line, but in general still lag behind the more traditional \ac{ann} solution in terms of accuracy \cite{Struye2019}. As understanding of learning processes in the brain further evolves, they may lead to better and more efficient learning techniques that can be implemented on ultra low power sensor nodes.

\end{itemize}


\section{Wireless for Neurosciences}
\label{sec:WforN}
Up to now, we have discussed different ways in which neurosciences would become part of the future wireless communication systems via \ac{btc}, specially under \ac{6g}.
In this section, we will describe how wireless communication theory and \ac{6g} systems can support future research and technological development in neurosciences.

\subsection{\acs{6g} performance for neurosciences}

While the plurality of applications are waiting for the real development of wireless \acp{bmi}, we must gather an initial assessment of the existing metrics, or new metrics, that allow the understanding of what is required in \ac{6g} for Brains to Wireless infrastructure connections. This initial analysis is made based on the recent breakthroughs in 5G and Beyond 5G research, which is the cornerstone of \ac{6g}, as well as recent engineering advancements in neural interfaces, which are the central pieces of \acp{bmi}. The key vision is to reach fine-granularity of brain functions from both sensing and actuation capabilities from integration with \ac{6g}. Then the desired performance of \ac{6g} is draw upon the ability delivering enough performance that maintain the well functioning of future \acp{bmi} for a long time with security and safety for users.

\subsubsection{Data rate} The naive estimation of whole brain recording demand is about 100 Gbits/s, which is not supported by existing and near-deployment 5G infrastructures. However, in the context of individual connections, this is a considerable demand for \ac{6g} which is currently not being considered due to the lack of popularity of \acp{bmi}. This estimation was also naively performed because it does not consider the current and future technology for \ac{bmi}, which surely can rise this number up as more and more techniques are capable to obtain not only electrophysiological neuron signals, as well as signals from other cell types in the brain and lastly other types of information such as biomarkers. On top of that, this naive estimation is also based on standard sampling rate of neural signal acquisition (1kHz), which surely varies between technology and recording strategies. The needs for increased data rate in \ac{6g} must deal with all aforementioned information, even though it needs more investigation about the real data rate requirements of \ac{bmi}. By looking at more spectrum resources, one must keep in mind that \ac{bmi} is one of the multiple applications that \ac{6g} systems must accommodate. Together with multimedia, gaming, e-health applications and more, \ac{bmi} can along increase the burden on future network generations for more data rate requirements than previously expected.


\subsubsection{Reliability} \acp{bmi} as a technology can open a wide-variety of applications sensitive to network disruption. As one example, remote treatment of epileptic patients will require a constant usage of the \ac{bmi} for detecting random seizure events as well as actuating upon the disease using current stimulation techniques also driven by the \ac{bmi} solution. The implications of network disruption in this case is above from conventional application delays or stream interruptions that it is commonly found in conventional networks. In this context, the disease control mechanisms that are based on \ac{bmi} solution could be disrupted in a way that either can start unpleasant symptoms for patients, as well as not supporting advance signal processing techniques for temporal variant data that support diagnosis systems. Based on the assumption of \acp{bmi} actively being used in small cells, that means not only that high frequencies must be managed to provide reliable connections that are not interfered by obstacles and environmental molecular effects such as water vapors. These phenomena are known as the biggest challenges in small cell for beyond-5G nowadays, where intelligent reflecting surfaces are being currently the best choice for provide highly reliable connections. However, this technology is far away from being mature to guarantee high levels of network reliability based on the primary focus on physical mechanisms of beam-steering of high frequencies as opposed to the study of network resilience, which must be the next step of the research in this topic. The importance of network reliability brings again the focus to solutions that maintain a constant data rate to certain applications, where conventional network solutions must be upgraded in \ac{6g}.

\subsubsection{Energy management} Wireless \acp{bmi} based on brain implants will most likely operate on a different wireless media than \ac{6g} wireless infrastructure. For example, while RF is an option for wireless \acp{bmi}, high frequencies are unlikely to be used due to signal absorption and scattering due to the tissue and skull high water molecule profile. However, the two better fitted options are magnetic-induction system as well as ultrasound. Their differences are highlighted by their performance profile, while magnetic-induction is better for implant data rate, ultrasound system enable deepness of implantation. The major challenge here is that \ac{6g} is most likely to operate around the sub-Terahertz bands, which means that constant frequency conversion is required in order to provide integration of wireless \acp{bmi} to \ac{6g} infrastructure. Since frequency homogenisation is not an option, it can be easily foreseen that \acp{bmi} must have short-term memory strategies that supports the frequency translations without loss of data. The issue here then is that both frequency conversion techniques and memory as energy costly, that aids the concerns for the \acp{bmi} constant usage for chronic patients, or other applications such as streaming or gaming. Energy management solutions must emerge to not only look at the device level, but at the network level, which can both work together using advanced protocols or virtual infrastructures that enables efficient and data lossless connections with \acp{bmi}. 

\subsubsection{Latency} Today's communication infrastructure is guided by techniques that provide massive ultra reliable and low latency communication. This shall not change for communication with \acp{bmi}. The importance of these strategies is directly linked with the future of \acp{bmi} and its success, since the main goal is to allow constant daily usage for patients and general users. The radical societal change from \ac{bmi} will only happen when we are capable of using this technology integrated into daily activity, either to support it or to enhance it. In \ac{6g}, it is promised the idea of massive sensing, which fits to the future \ac{bmi} technology which envision hundreds or thousand of nano-scale devices that interface with neuronal cells. The information in that scale, i.e. the \ac{lfp}, enable cellular rich information that is now used to make precise predictions of disease states and trigger events. In addition to that, the idea of massive stimulation can also be performed, where these several devices will act on the neural tissue for stimulating whole or parts of a population of cells. Latency here is crucial in order to operate these function remotely while maintaining the safety and security of each user. At the same time, this needs to be perfectly modeled and tackled in a possible way that allows scalability. Scalable \acp{bmi} are practically non-existent, and \ac{6g} might as well be the technology needed to open these doors.



\subsection{Internet-of-Bio-Nano-Things}

Another key promising application is the \ac{iobnt} \cite{akyildiz2015internet}. 
The \ac{iobnt} can aid the diversity of \acp{bmi} and their types by now interacting with the brain using molecules, peptides, and, overall, molecular structures \cite{balasubramaniam2018wireless}. As it stands, no molecules are used to convey synthetic information of any type, which means that a whole biodiversity of information is being underutilized as opposed to enhanced means of communication between implantable devices and brain tissue. The research area of molecular communications, promotes the usage of molecules as carriers for interactions between implantable-implantable and between implantable-biological systems \cite{akyildiz2008nanonetworks}. Increased biocompatibility is therefore reached when understanding and using molecules that are currently being used in biological system, now with the purpose of controllable biological communication \cite{barros2018feed}. This infrastructure is envisioned to bridge to the internet by means of synthetic biology and advanced nanotechnology, where electromagnetic-molecular signal translation is performed towards remote digital control of internal cellular process of either Eukaryote and prokaryotic cells.

The diversity of molecules inside a human body is assumed to be huge, and therefore the means of translating molecular information between tissues is assumed to be of great importance even with the limited investigation by the community so far. The idea for the future technology is that there are internal synthetic cells capable of converging molecular information from different types of tissues and vise-versa in order to support the idea of biomolecular intrabody networks \cite{barros2015comparative}. Therefore, implantables or even bionanomachines sited in different tissue can communicate with each other without the need of pre-determined molecular coherence, which can empower flexibility and performance of these systems. In the edge of these networks the molecular information is translated to electromagnetic information by biocyber interfaces, that are also capable of translating the opposite case\cite{chude2016biologically}.

Inside the brain implantables of bionanomachine devices have the main purpose of influencing the brain activity by manipulating the Ionic channels that are understood to be a major part of the information propagation in the brain. There are a variety of molecules in the micro-scales of the brain, including calcium, potassium and sodium. Neurotransmitters and gliotransmitters are ions that regulate the information propagation inside the synaptic channel between neurons. These molecules have been studied and analyzed by many decades and are controlled for purpose of treatment of many neurodegeneration diseases. Brian-machine interfaces for molecular interactions have a huge impact in the future of the neurodegeneration diseases. The levels of control that can be reached from digital systems can be tremendously beneficial to the always considered chaotic systems such as biological systems in the brain \cite{barros2017ca2+}. The main challenge is that the major biological properties of the brain are well understood before being considered as control variables, which is a time-consuming effort that has to focus on neuroscience efforts that are being developed through many decades.

However, there are existing works that demonstrate the idea of \ac{iobnt} prevailing though existing disease challenges together with biotechnology as well as future oncology efforts. The EU-H2020-FET Gladiator project utilizes a hybrid neural interface that is implanted into the brains of patients suffering of Glioblastoma brain cancer with the main goals of utilizing the modulation of drug propagation in the brain that maximizes drug efficacy while minimizing drug side effects \cite{veletic2019molecular}. For that, wireless external signals control these hybrid neural interfaces to produce molecules that contain multiple drug molecules inside them, called exosomes. These exosomes are the drug mediation agent that ultimately dictate how and when the brain cancer is being decreased from this novel treatment. The novel paradigm of molecular communication is being utilize to characterize the data rate and capacity of exososome-based communication systems between the hybrid interface and the brain cancer. Here the channel is understood to be the extracellular brain space that the exosomes can propagate through a biased random motion. The many brain cells create the tight spaces where the exosomes propagate where there is enough brain fluid to drive movement, called brain parechyma. Researchers are now focusing on developing both theoretical and in-vitro models that demonstrate the above-mentioned system that can radically change the existing state-of-the-art of cancer oncology treatment methods.

\subsection{Wireless-based brain-machine interfaces}

A more direct application of 6G technology that neurosciences would benefit would be a new generation of \acp{bmi}.
\acp{bmi} have been used to alleviate motor deficits but also as a tool to characterize neural correlates of behavior \cite{WOLPAW2002767}. In this sense, tethered neural recording systems are a major limitation because it hinders natural and social behavioral interactions. Most notably in the late 2000s, novel wireless recording technology became available which can simultaneously sample several hundred neurons from different brain regions (see \cite{Hong2019} for a review on recording technology).

Schwarz and collaborators \cite{Schwarz2014} developed a bidirectional wireless system capable of implementing part of the signal processing pipeline at the headstage and transceivers attached to the animal's head. Four transceivers were used, each transceiver was connected to 128 recording channels sampled at 31.25 kHz per channel, consuming 2 mW per channel, with a total of 48-Mbps aggregate rate of data acquisition, and an optimal operating range of 3m. The device is reported to be able to continuously operate over 30h. Implanted in a monkey, authors were able to record 494 neurons from 4 different brain regions. The animal successfully performed established \ac{bmi} tasks wirelessly \cite{Rajangam2016}, thus confirming its suitability for studying natural, social interactions and complex movement behaviors.

One major problem is common to invasive recording systems: implants inevitably lesion brain tissue, causing inflammation and limiting the sampling of deeper brain regions. To alleviate this problem, wireless sub-millimeter scale devices are being developed that can both record and stimulate neural activity. Ghanbari and collaborators \cite{Ghanbari2019} describe an ultrasonically powered neural recording implant, with simultaneous power-up and communication, that can achieve over 35 kbps/mote equivalent uplink data rate. Thus, in principle, this device could operate as part of a wireless \ac{bmi}.

In terms of non-invasive wireless \acp{bmi}, EEG is arguably the most common recording strategy. By avoiding surgical procedures, EEG has been widely used in human \acp{bmi}. However, EEG signals reflect the activity of millions of neurons from the surface of the brain, thus hindering decoding performance which leads to a limited set of motor commands that can be extracted in non-invasive \acp{bmi}. Common commercial wireless devices range from 8 to 64 channels, with sampling frequencies of up to 1 kHz and a few dozen meters of transmission \cite{Lin2010}. Nevertheless, modern hardware and computational intelligence methods, such as flexible electronics and deep learning, respectively, have been shown to boost wireless EEG-based \ac{bmi} performance reaching up to 122 bits per minute of information transfer rate \cite{Mahmood2019}.

\subsection{Brain as a complex system with chaotic communications}
Communications and information theory tools can also provide interesting analytical  approaches to assess the behavior and the fundamental limits of neural communications, which is chaotic by nature.
%
In physics, chaos refers to states that lie between order and randomness.
Chaos brings rich dynamics that are governed by deterministic processes, capable of inducing non-trivial patterns and behavior.
In the early days of the chaos signal processing, enabled by new algorithms for attractor reconstruction and Lyapunov exponent calculation in the nineties, EEG signals were one of the first major targets for analysis \cite{fell1993deterministic}. 
The announcement of ``chaos in the brain'' was followed by the discussions of complexity in the brain, measured by different entropic measures. 
Schizophrenia is a decay of complexity, either on the side of randomness (infinite entropy), or order (zero entropy) \cite{yang2015decreased}. 
Furthermore, the effects of drugs like LSD are connected to periodic behavior of the brain, again losing chaoticity and decreasing complexity \cite{dubois2011visual}.
In a complex network formed by neurons, different parts can exhibit different behavior and yet be inseparate: this is an example of chimera states, studied in theory and found in nature \cite{wei2018nonstationary}.

Even though the effort to discover organization in nature had its origins in randomness, it was realized that measures of randomness do not capture the property of organization \cite{feldman2008organization}. This led to the development of measures that capture a system's complexity - organization, structure, memory, symmetry, and pattern. This led to the development of measures that capture a system's complexity - organization, structure, memory, symmetry, and pattern. Complexity allows us to quantify the hidden micro-level relationships between system parts that result in the system properties obtainable on the macro-level. The authors of \cite{tononi1998complexity,bialek2001complexity} argue that a complex system lives in-between a random and a completely regular system, leading to the conclusion that a lot of the so called complexity metrics (e.g.~Kolmogorov complexity in algorithmic information theory, dimensional complexity in neurobiology) do not measure ``true complexity`` because they do not attain small values for both random and regular systems. Random systems have no structure at any level, which results in high entropy and low complexity. On the other hand, regular systems exhibit low entropy and low complexity due to the repetition of structures on multiple levels. Therefore, it is obvious that complexity and entropy are two distinct quantities. As highlighted in \cite{feldman2003}, entropy captures the disorder and inhomogeneity rather than the correlation and structure of a system. Therefore, even though we believe that we can recognize complexity when we see it, complexity is an attribute that is often without any conceptual clarity or quantification. 

The difference between completely regular, random and complex systems as described in \cite{tononi1998complexity} has been addressed earlier by \cite{weaver1948}. However, the author of \cite{weaver1948} refers to these concepts as simple, disorganized complex and organized complex systems respectively. He also highlights that scientists were finding ways to model simple and disorganized complex systems, while the problem of organized complexity has not been tackled properly before. He also highlights that the challenges related to organized complexity are related to two main issues: (1) macroscopic predictions for organized complex systems are impossible due to the interactions between a large number of dynamic variables on the microscopic level; (2) the emergent, self-organizing whole created by the system parts is dependent and comprised of multiple causal models. Hence, organized complex systems can not be simply modeled by mathematical formulas, qualitative description or statistics. Instead, new approaches to science, along with new modeling methods and studies of the relationship between complexity metrics and the emergent behavior of those systems are needed to connect the system organization on the micro and macro level.

According to \cite{Lloyd2001}, three main questions (1. How hard is it to describe? 2. How hard is it to create? 3. What is its degree of organization?) have to be answered before quantifying the complexity of a system. These questions can then be used to classify different measures of complexity. Measures like: information, entropy, code length, etc. allow us to quantify the degree of difficulty involved in describing the system - typically quantified in bits. The difficulty of creation, which is typically quantified by time and energy, includes measures like: computational complexity, cost, crypticity, etc. The degree of organization includes measures like: excess entropy, hierarchical complexity, tree subgraph diversity, correlation, mutual information. Not all of the mentioned metrics are a measure of complexity per se, but all of them can be used to define different complexity metrics. Some examples include: 
\begin{itemize}
    \item Excess entropy measures the degree of organization of a system. It is an information-theoretic measure of complexity. As shown in \cite{feldman2003}, it measures the amount of apparent randomness at the micro-level that is ``explained away`` by considering correlations over larger and larger blocks. Completely random and fully structured system configurations exhibit low excess entropy, whereas structures with a certain level of organization without pattern repetition on different system scales exhibit high excess entropy. 
    \item Neural complexity was introduced by the authors of \cite{Tononi1994}. It measures the amount and heterogeneity of statistical correlations within a neural system in terms of the mutual information between subsets of its units. In other words, it captures the interplay between global integration and functional segregation, resulting in high complexity for a system in which functional segregation coexists with integration and low complexity when the components of a system are either completely independent (segregated) or completely dependent (integrated). 
    \item Matching complexity, which was introduced in \cite{tononi1996complexity}, builds on the work in \cite{Tononi1994} and reflects the change in \textit{neural complexity} that occurs after a neural system receives signals from the environment. It measures how well the ensemble of intrinsic correlations within a neural system fits the statistical structure of the sensory input. \textit{Matching complexity} is low when the intrinsic connectivity of a simulated cortical area is randomly organized and high when the intrinsic connectivity is modified so as to differently amplify those intrinsic correlations that happen to be enhanced by sensory input. 
    \item Statistical measure of complexity is defined as the product of two probabilistic measures: disequilibrium and entropy \cite{Lopez-Ruiz1995}. Disequilibrium gives an idea of the probabilistic hierarchy of a system, and it is high for highly regular systems and low for disordered/random systems. On the other hand, entropy is low for ordered systems and high for disordered systems. The product of these two values provides a complexity measure that is high for systems that are in-between ordered and disordered systems. 
    \item Self-dissimilarity measures the amount of extra information using a maximum entropy inference of the pattern at one scale, based on the provided pattern on another scale \cite{wolpert1997self}. It characterizes a system's complexity in terms of how the inferences about the whole system differ from one another as one varies the information-gathering space.
\end{itemize}

The account of complexity metrics does not stop here. Authors like \cite{shiryayev2013selected, chaitin1990information, lloyd1988complexity, gell2002complexity, ribeiro2012, rosso2007extracting} have taken different approaches by relating complexity to mathematical formulations of thermodynamics or statistical entropy and information.
Those metrics are mostly employed in the existing literature to study the complexity of the user behavior and the impact that external factors have on the network.

However, the communication network itself can be analyzed as a complex system by focusing on its organizational structure that affects the execution of network functions giving its complexity. 
Authors of \cite{Wang2009,Onnela2007,hidalgo2008,candia2008,Beigy2010,Macaluso2014,Macaluso2016} apply complex systems principles to understand different phenomena in communication networks. In \cite{Wang2009,Onnela2007}, the authors analyze complex phenomena in telecommunication networks that resulted from complex user behaviors (e.g.~spreading patterns of mobile viruses and connection strengths between nodes in a social network). The authors of \cite{hidalgo2008} study the correlation between the structure of a mobile phone network and the persistence of its links, highlighting that persistent links tend to be reciprocal and are more common for nodes with low degree and high clustering. The authors of \cite{candia2008} study the human dynamics from mobile phone records. This studies reveal the mean collective behavior at large scale and focus on the study of anomalies. These studies of human dynamics have direct implications on spreading phenomena in networks. However, unlike the work in these papers that focuses on the complexity of the user behavior, the work in \cite{Macaluso2014, Macaluso2016, dzaferagic2018functional, dzaferagic2017WSN, pattanayak2018VANET} examines the complexity of the operation of communication networks themselves. The authors of \cite{Macaluso2014,Macaluso2016} analyze the complexity of outcomes of a self-organizing frequency allocation algorithm. In \cite{Macaluso2016}, they analyzed the relationship between robustness and complexity of the outcome. More precisely, the work in \cite{Macaluso2014, Macaluso2016} does not focus on the interactions between network nodes. They rather focus on the patterns that emerge from those interactions, i.e.~they study the patterns after the self-organizing allocation algorithm converges. On the other hand, the authors of \cite{dzaferagic2018functional} study the interaction between the network nodes during the execution of network functions and its effects on the overall network operation. The authors of \cite{dzaferagic2017WSN, pattanayak2018VANET} apply the theoretical framework introduced in \cite{dzaferagic2018functional} to study the trade-off between energy efficiency and scalability in \ac{wsn} and the affect that the complexity of the underlying structure has on the probability of collision and probability of correct packet detection in vehicular ad-hoc networks. 
In this sense, the brain can be also studied as a complex communication network associated with structure and function, and evaluated with information-theoretical-inspired metrics and distributed communication systems performance indicators. This knowledge translation from complex networks to brain network can shine a light on organization and structures of the brain that are currently unknown.

\subsection{Chaos-based communications}

The scientific understanding of chaotic systems, as described before, also opened the opportunity to design chaos-based communication systems.
Somehow, the precursor of this idea was Shannon himself in his seminal work, in which was demonstrate that a noise-like signal with a waveform of maximal entropy results on optimized channel capacity in communications~\cite{Shannon659497}. 
A fundamental advance for the practical implementation of chaos-based systems communications only occurred at 1980 with the Chua's work~\cite{Chua1084745}. His experimental circuit, which is called the Chua's circuit, presented chaotic behavior. 
Roughly speaking, chaotic behavior refers to a system whose dynamical is very sensitive to the initial conditions. 
In other words, given a set of the initial conditions for the chaotic system, we can always find others initial conditions arbitrarily nearby that lead to drastic changes in the eventual behavior of the system~\cite{bookHirsch}.

Basically, a vast number,  theoretically infinite,  of signals decorrelated can be generated with a small variation in the initial conditions of the system. 
This makes it possible to use chaotic signals for encrypting messages and multi-user spread-spectrum modulation schemes~\cite{LinChaosEncryp,bookLau,KurianDS/CDMA}. Moreover, in some specific case, chaotic-base modulations have provided the same advantages as conventional
spread-spectrum modulations~\cite{Lynnyk,Wagemakers_2008}.
For example,  the performance of a system operating with \ac{dcsk} modulation over a severe two-ray Rayleigh fading channel is best than with \ac{bpsk} modulation, on the other hand, over a negligible fading channel, the modulation \ac{bpsk}-based present best performance than \ac{dcsk}-based~\cite{Xia}.

It is worth mentioning that the \ac{dcsk} modulation showed the same performance under both hypotheses: severe or negligible two-ray Rayleigh fading channel. 
Other examples are jamming resistance along with \ac{lpi}~\cite{YuandYao} and secure communications~\cite{Lynnyk, Wagemakers_2008}.

Chaos-based communication can use coherent or non-coherent detection. In the first case, the receiver must be able to generate a replica of the transmitted chaotic signal~\cite{Dedieu,Parlitz,Kolumban}.
In the second case, there is not the necessity of regeneration of the signal chaotic in the receiver side~\cite{bookLau}. 
Based on this important advantage (simplified receiver), the non-coherent chaos-based communication has gained primary attention over the years.
Observe in Table \ref{table: Modulations classes} (refer to ~\cite{Kaddoum7478568}) there is a vast list of modulation schemes with non-coherent detection, whereas the number of coherent modulation schemes is limited to five.

Adding a chaotic oscillator to the information leads to a spectral spreading process. In a simplified way, we can say that chaotic oscillator is equivalent to \ac{pn} sequences in conventional binary spreading.
From a security perspective, chaotic-based communications are more robust than the \ac{pn} sequences. 
The main reason is the susceptibility of \ac{pn} sequences to reconstruction by linear regression~\cite{bookPeterson}.

Communication-based on chaotic oscillators continues to be a target of studies, and its practical implementation has proved to be challenging. Recently, some efforts have been made to demonstrate in practice the viability of chaotic communication. More specifically, mitigating the main drawback of a chaotic-based system: the synchronization of chaotic oscillators. 
As we will discuss, establishing the synchronism between two chaotic systems is not a trivial process. Because of this characteristic, one of the most promising applications for chaotic systems is in the area of communication security~\cite{Wagemakers_2008}.
%
%
\begin{table}[t!]
	\centering
	\caption{Modulations classes. Adapted from~\cite{Kaddoum7478568}}	
	\begin{tabular}{cc} 	
		\textbf{Detection} & \textbf{Modulation}\\ \hline\hline
		
		& Chaos Shift Key \\
		& Symmetric CSK  \\
		Coherent & Chaos based CDMA \\
		& Quantized chaos based CDMA \\
		& Chaotic Symbolic dynamics \\
		\hline
		& Differential CSK \\
		& FM DCSK \\
		& Ergodic CSK \\
		& Quadrature CSK \\
		& M-DCSK \\
		& DCSK-WC \\
		& PMA-DCSK \\
		Non-coherent & HE-DCSK \\
		& CS-DCSK \\
		& UWB-DCSK \\
		& PS-DCSK \\
		& RM-DCSK \\
		& MC-DCSK \\
		& DDCSK \\
		& I-DCSK \\
		\hline
	\end{tabular}
	\label{table: Modulations classes}
\end{table}

The synchronization process in the receiver's chaotic oscillator is strongly linked to a perfect match of the initial condition parameter. 
However, assuming that the receiver knows these parameters, the synchronization of the chaotic oscillator in the receiver becomes achievable. 
Several synchronization approaches have been recommended in the literature for chaos-based communication systems. 
These approaches fall into two categories: chaos synchronization techniques~\cite{PhysRevLett.64.821} and conventional synchronization approaches applied to chaos-based communication systems~\cite{JovicCDMA}. 
The first category resembles master-slave systems, more explicitly, a chaotic oscillator (slave) is forced to follow a reference signal (master) in order to achieve synchronization. 
As we will discuss later, it is not always clear which system is the master and which assumes the role of the slave. 
The second category, as the name suggests, is base on conventional methods of synchronization such as phase and sampling synchronization to chaos-based coherent communication systems.

The concept of master-slave synchronization, as to the coupling between chaotic systems, can be summarized in four types~\cite{Kaddoum7478568}:
\begin{itemize}
	\item \textit{Directional Driving}: One chaotic system is used as the source of driving transmitting one or more driving signals to the other system, there is no mutual interaction between the systems involved~\cite{Stojanovski1996}.
	\item  \textit{Bidirectional Driving}: Two chaotic systems are coupled and mutually driven with each other~\cite{Zhou2007}.
	\item  \textit{External Driving}: External signal driving the chaotic systems present in a network to establish synchronization. It can be considered an extension of the Directional Driving for a network~\cite{ZLi2006}.
	\item  \textit{Network coupling}: Many chaotic systems are coupled with each other. This named a complex dynamic network. It can be considered an extension of the Bidirectional Driving for a network~\cite{Pikovsky1997}.
\end{itemize}
Coupling is particularly relevant when dealing with communications based on chaotic systems. In practical terms, we can interpret coupling through the communication channel. And, in today's communications context, systems are commonly designed to operate with full-duplex communications, enabling multi-directional coupling.

As long as the parameters of the chaotic circuits are adequately tuned, it is possible to obtain synchronous behavior in two identical chaotic circuits with bidirectionally coupled or coupled through another chaotic circuit~\cite{Wagemakers_2007,Wagemakers_2008}. In lag synchronization, for example, as the name suggests, the receiver follows the evolution of the transmitter with a delay due ta a parameter mismatch. 
If there is a time-lapse between the output of the systems synchronized, we call it achronal synchronization. 
On the other hand, systems highly synchronized, without any delay between outputs, despite the time lost in the transmission line, are isochronal synchronization.

In all cases where the unidirectional coupling is considered, the leader and follower role (master and slave) can be identified. 
The system that sends the samples of itself oscillator is the leader (master) and the other one is the follower (slave). 
In the unidirectional coupling, this role appears to be easily distinguished. 
However, for bidirectional coupling, it is not clear who is driving and which systems are being driven. 
Because there is a mutual influence between the systems involved~\cite{Wagemakers_2007,Wagemakers_2008}.

\begin{figure}[t!]
\includegraphics[width=0.92\linewidth]{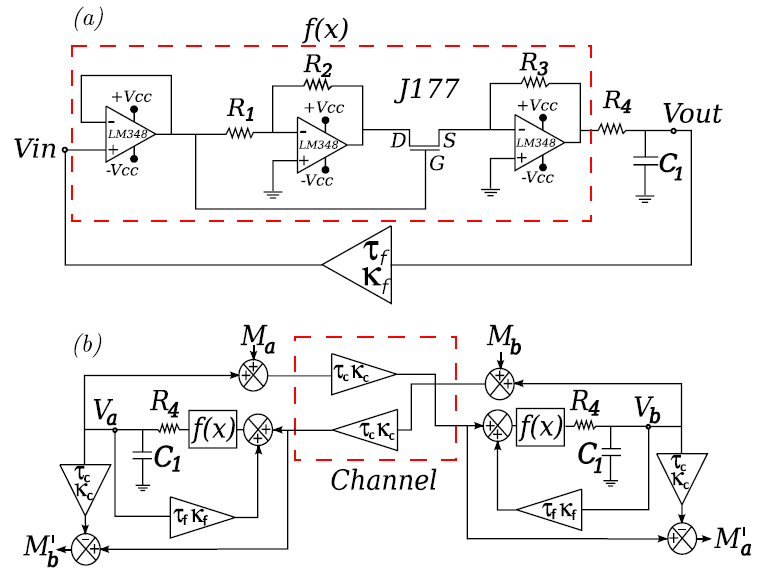}
	\caption{In (a) a single Mackey-Glass circuit is represented. The circuit consists of feedback with gain $ \kappa_f $ and delay $ \tau_f $, a nonlinear function $ f(x) $ and an RC filter. In (b) we plot the schematic setup of the transmission of a message ($ M_a $ or $ M_b $) with chaotic masking.}
	\label{fig: MackeyGlassCircuit}
\end{figure}

As an example of a chaotic oscillator, \figurename{\,\ref{fig: MackeyGlassCircuit}} outlines the Mackey-Glass electronic circuits~\cite{Wagemakers_2008}. There are some other circuits with chaotic behavior, such as the Chua circuit~\cite{Chua1084745}. 

Although much of the literature shows experimental results, in this paper we propose a nonlinear function $ f (x) $, given by 
\begin{equation}\label{eq: FunMackeyGlass}
f(x) = G \frac{\alpha  \mu ^{\mu } x^{\alpha  \mu -1}}{\Gamma (\mu ) \hat{x}^{\alpha  \mu }} \exp \left(-\mu  \left(\frac{x}{\hat{x}}\right)^{\alpha }\right),
\end{equation}
where $ \Gamma(\cdot) $ is the Gamma function
and $ G $, $ \alpha $, $ \mu $, and $ \hat{x} $ are parameters that must be properly adjusted. 
This function is equivalent to the block of amp-ops and FET (J177) in \figurename{\,\ref{fig: MackeyGlassCircuit}}. We believe that this is of great value for reproducing and understanding the transmission process based on chaotic oscillators. 

\begin{figure}[t!]
	\centering{\includegraphics[width=0.92\linewidth]{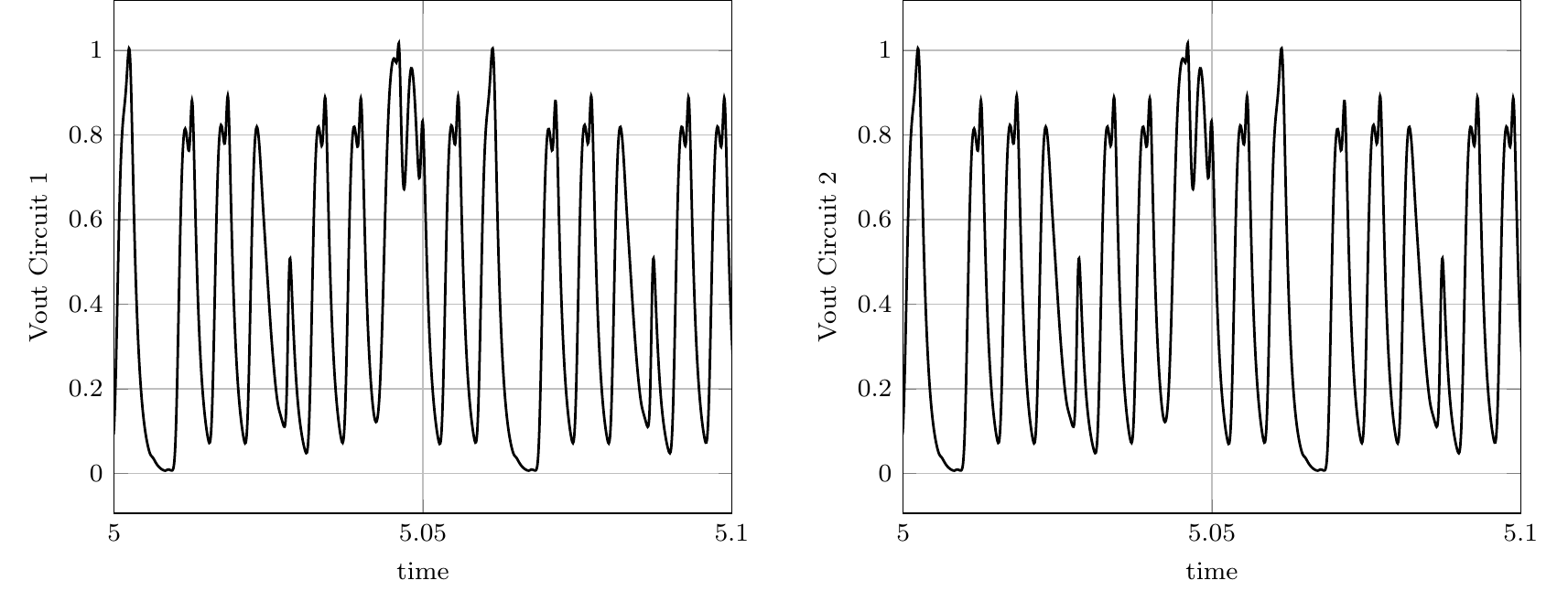}}
	\caption{Sample of the output signal of two Mackey-Glass circuits coupled and with perfectly matched parameters.}
	\label{fig: VoutCorrelat}
\end{figure}

\begin{figure}[t!]
	\centering{\includegraphics[width=0.92\linewidth]{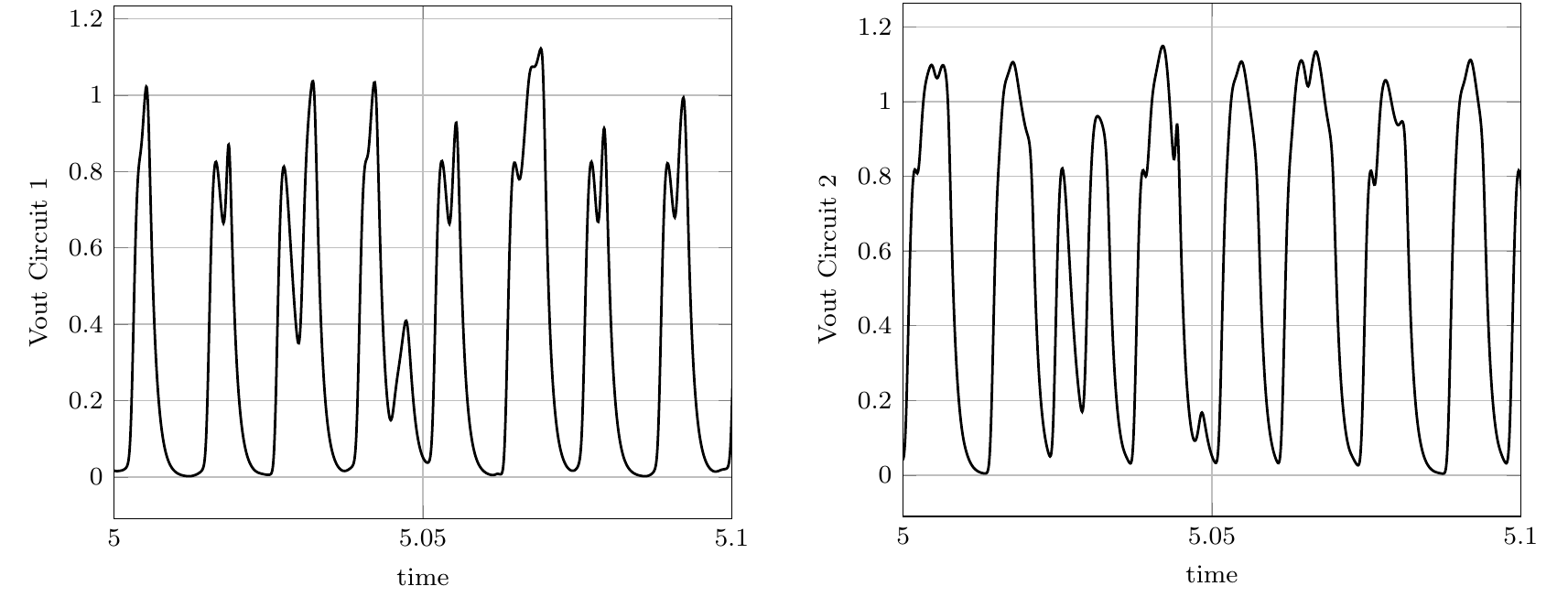}}
	\caption{Sample of the output signal of two Mackey-Glass circuits coupled and with mismatched parameters.}
	\label{fig: VoutDe-Correlat}
\end{figure}

\figurename{\,\ref{fig: VoutCorrelat}} plots a sample of the outputs of two mutually coupled oscillators with their respective perfectly matched parameters.
The values of the parameters used in the simulation were $ G=0.7 $, $ \alpha=2 $, $ \mu=1 $, $ \hat{x} = 0.4 $, $ \tau_c = \tau_f=0.018 $, $ \kappa_c = 1 $, $ \kappa_f=0.4 $, $ \text{R}_4 = 1k\Omega $ and $ \text{C}_1 = 1\mu\text{F} $.  
The signals were generated through simulation using \eqref{eq: FunMackeyGlass}. 
Observe the perfect synchronization in the signals present in \figurename{\,\ref{fig: VoutCorrelat}}, while in \figurename{\,\ref{fig: VoutDe-Correlat}}, the parameter $ \tau_f $ was purposely mismatched $\tau_c \neq\tau_f = 0.015 $, producing uncorrelated signals at the outputs of the coupled oscillators. Certainly, the perfect match of parameters is the great challenge of communication-based on chaotic circuits.

\section{Case study: Brain-controlled vehicles}
\label{sec:case}
In this section, we will present the state-of-the-art of one particular application named \ac{bcv} and how we foresee its development based on \ac{6g}.
Note that \ac{bcv} is an interesting application that in fact involves the two groups of contributions described here. At the same time, it is relevant for neuroscientists studying motor control, while it is also an engineering problem since motor intentions need to be reliably mapped into actions in vehicle control (which we expect to be a particular case in \ac{6g} systems).

\begin{figure*}[t]
  \begin{center}
  \includegraphics[width=0.65\textwidth]{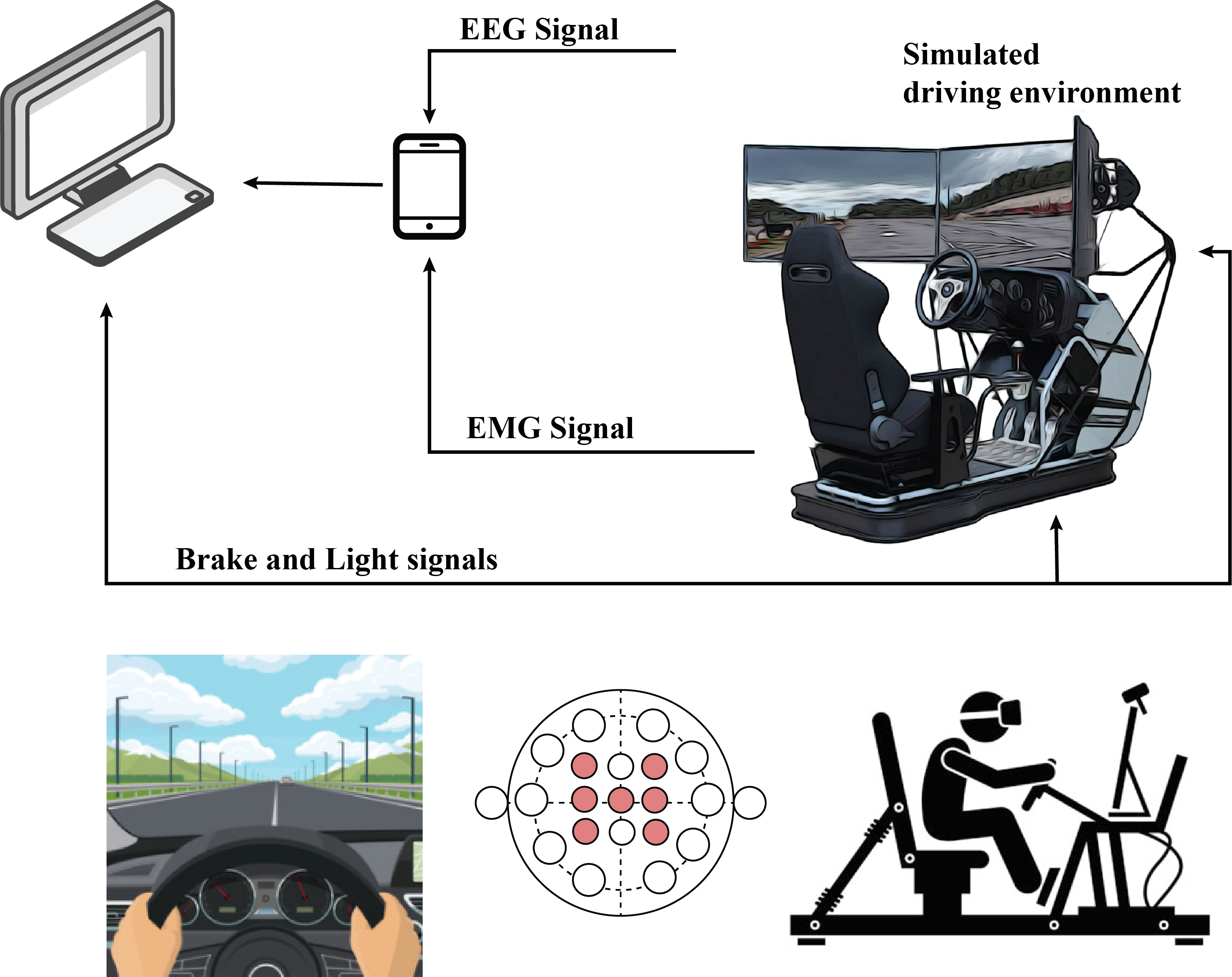}
  \caption{A sample driving training framework including computer simulators, EEG, EMG, and a real-time control system. Adapted from \cite{hernandez2018eeg}.}
  \label{simulator}
  \end{center}
\end{figure*}

\subsection{State-of-the-art}

The field of \ac{bcv} with EEG-based \ac{bmi} has experienced a steadily growth since 2010, usually focusing on applications to support disabled patients.
In addition to the already discussed challenges of \ac{bmi} in relation to developing effective algorithms for feature extraction and classification, current \ac{bmi} hardware has well known limitations concerning communication range and speed.
Connecting the \ac{bmi} device with the computer and sending commands to the vehicle is constrained by the communication link.
When the communication link is wireless, the most usual approach is to use WiFi or Bluetooth.

Despite those challenges, several studies have been published that demonstrate the feasibility of \ac{bcv}, mainly studying algorithms for vehicle control.
We can cite, for instance, controlling a vehicle in four main directions \cite{lu2019combined,hekmatmanesh2019optimized}, methods for obstacle avoidance \cite{hernandez2018eeg,zhuang2019ensemble}, and hand brake assistance in emergency situations \cite{bi2018novel,teng2017eeg}, all based on driver’s intention. 
The above-mentioned topics are explored in different applications such as controlling (i) vehicle simulators, (ii) virtual reality vehicles, (iii) vehicles in video games, (iv) quadcopters, (v) drones, (vi) helicopter, and (vii) fixed wings aircrafts. A general simulator-based procedure for training a participant is shown in Fig. \ref{simulator}.

In a seminal work, Haufe et al. \cite{haufe2011eeg} implemented an assistant brake system in emergency cases for BVC applications based on EEG and EMG signals. The experiment is then tested on a simulated vehicle. The algorithm map brain activity patterns related to emergency brakes in a simulated graphical racing car task.
In a similar line, Kim et al. \cite{kim2014detection} attempted to detect the driver’s emergency brake intention in different situations for a simulated vehicle based on EEG and EMG signals.
This method was further improved in \cite{haufe2014electrophysiology}, also considering a real-time experimental task.
In Gohring et al. \cite{gohring2013semi}, a semi-autonomous vehicle is implemented with different external sensors, camera, and then controlled using EEG-based brain activity patterns. For controlling the vehicle, two different scenarios, obstacle avoidance and braking and steering, were used. 

In a series of studies by Bi et al. \cite{bi2013using,bi2013head,fan2014brain,bi2016queuing,bi2018novel,lu2019model}, different approaches to identify and predict the driver's intention for going forward, turning left and right, as well as emergency braking, were studied. 
The development of AI-based learning methods are leading to improvements in those tasks, as reported in \cite{hekmatmanesh2020combination,hekmatmanesh2019combination,hekmatmanesh2019optimized,hekmatmanesh2019optimizing,hekmatmanesh2019eeg,hekmatmanesh2018common,hekmatmanesh2019investigation,hekmatmanesh2017sleep}.
Similar research has been also carried out to study \ac{bcv} for aerial vehicles, as in \cite{zafar2018drone,khan2017hybrid,shi2015brain}. A major challenge is in separating from brain signals features that relate to vehicle control from those that are not related. The second part for enhancing the results is developing or modifying the existing classifiers into highly accurate multi-classifier, such as deep believe learning algorithm. The third limitation is the limited number of participants for training and testing of the algorithms. 

\subsection{Future vision: Neurosciences-6G converged applications}
Although successfully tested under different conditions, \ac{bcv} as designed today is not a scalable solution since this would require a wireless connection to support \ac{btc} with high coverage, availability, speed, and low latency to provide reliability and safety for the end-users.
As discussed before, despite of the great development of wireless communications (remarkably 5G), the existing solutions would not work today due to the stringent requirements of \ac{btc} (see Section \ref{sec:NforW}).
However, if the path indicated in this paper would be realized, a scalable \ac{bcv} would become feasible by using new generation of wireless-connected \ac{bmi} with 6G-connected high-density implants supported by \acp{irs} to enable \ac{btc}.
This would also be associated with the possibility of acquiring and processing more biosignals via IoTBNT, linked with \acp{isn} that could sample the environment in which the \ac{bcv} is moving.
The performance limits of those communicative brain devices could be derived from information- and communication-theoretical tools applied for chaotic and spiking systems, while new chaos-based waveforms for communication might be also developed.

As a rough example, we could imagine the following future scenario in 10-15 years from now.
Considering a \ac{bcv} system that would support the delivery of gift from Alice to Bob, who lives 15 kilometers away from her. 
Alice could sit in her armchair with a 6G-enabled \ac{bmi} that is synchronized with the \ac{bcv} to be used to deliver that specific good. 
Via \ac{bcv}-enabled \ac{ar}, the \ac{bcv} could be semi-autonomously controlled by hand movements and brain-signals.
The \ac{qope} to support this application should be guaranteed, considering that it requires not only \ac{btc} support but also other classes of communication. For example, the communication system must support the uploading of dynamic maps, associated with 3D holographic transmissions, and \ac{ar} integration with the video transmission from the \ac{bcv}. Finally, an alarm to communicate  Bob that his gift is arriving.

Clearly, this scenario could be extended and rethought, but it illustrates a potential future that we believe is technologically feasible given the state-of-the-art in wireless communications and neurosciences, as well as in biosignal processing and computer sciences. 
All in all, the convergence of those fields stimulated by \ac{6g} research agenda would make those kind of potential futures a reality for many applications related to future 6G-connected \acp{bmi}, as the \ac{bcv} example indicated. 

\section{Concluding remarks}
\label{sec:final} 
This tutorial paper provided an in-depth overview of an interdisciplinary research field at the intersection of neurosciences and wireless communications, as well as signal processing, control theory, and computer sciences.
We argue here that this convergence will take place in the coming \acl{6g}, which will support \ac{btc} considering not only its strict requirements but also the particularities of neural signals and brain communication.
By revisiting the available literature, we have classified the expected benefits of this joint research into two groups, as follows.
\textit{Neurosciences for wireless} focuses on how developments in neurosciences will enable new application in \ac{6g} based on \ac{btc}, as well as new architectures of sensor networks that build ``artificial brains''.
\textit{Wireless for neurosciences} focuses on how \ac{6g} would support new research and development in neurosciences, including novel 6G-enabled \acp{bmi}, and \ac{iobnt}, as well as information- and communication-theoretic ways of evaluating brain communications based on their chaotic nature.
We illustrated the potential benefits of this proposed research agenda by analyzing a brain-controlled vehicle application.

We expect this contribution will serve as a key reference for researchers from both domains to start building joint activities that are necessary to realize the vision indicated here.
The proposed discussions shall point towards a direction full of potential, from basic research to product development, but that can only be realized as a truly interdisciplinary task, similar to the path taken by neuromorphic computing \cite{strukov2019building}.

\bibliographystyle{IEEEtran}
\bibliography{walid.bib,IEEEabrv.bib,ref.bib}

\end{document}